\documentclass{ptephy_v1}

\usepackage{amsfonts,latexsym,eucal,color,graphicx,subfig,amssymb,amsmath,booktabs,multirow}
\newtheorem{Def}{Definition}[section]

\newtheorem{Lem}[Def]{Lemma}
\preprintnumber{RUP-15-31} 

\begin{document}
	
	\title{Spherical and nonspherical models 
		of primordial black hole formation: exact solutions}
	
	\author[1]{Tomohiro Harada}
	\affil[1]{Department of Physics, Rikkyo University, Toshima,
		Tokyo 171-8501, Japan \email{harada@rikkyo.ac.jp}}
	
	\author[2,3]{Sanjay Jhingan}
	\affil[2]{Centre for Theoretical Physics,
		Jamia Millia Islamia, New Delhi-110025, India}
	\affil[3]{iCLA, Yamanashi Gakuin University, 
		2-4-5, Sakaori, Kofu-City, Yamanashi,  400-8575, Japan}

\begin{abstract}
We construct spacetimes which provide spherical and nonspherical models
 of black hole formation in the flat Friedmann--Lemaitre--Robertson--Walker
 (FLRW) universe with the Lemaitre--Tolman--Bondi solution and the
 Szekeres quasispherical solution, respectively. These dust solutions
 may contain both shell-crossing and shell-focusing naked
 singularities. These singularities can be physically regarded as the
 breakdown of dust description, where	strong pressure gradient force
 plays a role.	We adopt the simultaneous big bang condition to extract
 a growing mode of adiabatic perturbation in the flat FLRW universe. If
 the density perturbation has a sufficiently homogeneous central
 region and a sufficiently sharp transition to the	background FLRW
 universe, its central shell-focusing singularity is globally covered.
 If the density concentration is {\it sufficiently large}, 
 no shell-crossing singularity appears and a black hole is formed. If the
 density concentration is {\it not sufficiently large},	a shell-crossing
 singularity appears. In this case, a large	dipole moment
 significantly advances shell-crossing singularities	and they tend to
 appear before the black hole formation. In contrast, a
 shell-crossing singularity unavoidably appears in the spherical and nonspherical evolution of cosmological voids.	The present analysis is general and applicable to cosmological nonlinear structure formation described by these dust solutions.
\end{abstract}
	
\subjectindex{E00, E01, E31}

\maketitle

\section{Introduction}

Black holes may have formed in the early Universe. These black holes are called primordial black holes (PBHs) 
\cite{Zeldovich:1967, Hawking:1971ei, Carr:1974nx}. Since PBHs can act as sources of Hawking radiation, strong gravitational fields and gravitational radiation, current observations can place a stringent upper limit on the abundance of PBHs 
\cite{Carr:2009jm}. Generally speaking, PBHs of mass $M$ were formed at
an epoch when the total mass enclosed within the Hubble radius was of
order $M$, if we neglect critical behavior in gravitational collapse,
mass loss due to Hawking evaporation, and mass gain due to
accretion. The abundance of PBHs of mass $M$ depends primarily on the
primordial fluctuations of mass $M$ and the state of the Universe at the
formation epoch. Since the Hawking radiation and its final outcome
depend on quantum gravity, PBHs can also provide observable phenomena of
quantum gravity, even if no positive signal has been detected. Thus,
PBHs can be seen as a probe into the early Universe, high energy
physics, and quantum gravity. In particular, since PBHs can convey
information about primordial fluctuations, the observational constraint
on the abundance of PBHs is complementary to the cosmic microwave background anisotropy observation.

To convert primordial fluctuations in a given cosmological scenario to
the abundance of PBHs, we need to understand the detailed process of PBH
formation. The process of PBH formation entails big bang, gravitational
collapse, apparent horizon, relativistic fluid, primordial fluctuations
and cosmological expansion. We need dynamical and inhomogeneous
solutions of the Einstein equation and the equations of motion for the
matter fields. The threshold of PBH formation in spherical symmetry was
analytically derived based on the Jeans criterion by
Carr~\cite{Carr:1975qj}, and has recently been refined by Harada, Yoo
and Kohri~\cite{Harada:2013epa}. The threshold of PBH formation in
spherical symmetry has also been studied by hydrodynamical simulations
based on numerical relativity, which was pioneered by Nadezhin et
al.~\cite{Nadezhin:1978}. For recent numerical studies, see Refs.~\cite{Musco:2012au,Nakama:2014xwa,Harada:2015yda,Bloomfield:2015ila} and references therein.

In summary, these studies support efficient PBH production for a soft equation of state. The effective equation of state becomes very soft if particles are very massive~\cite{Khlopov:1980mg} or if a scalar field oscillates in a quadratic potential, as in the ending phase of inflation~\cite{Suyama:2004mz, Suyama:2006sr}. A fluid with a very soft equation of state can be well approximated by dust or pressureless fluid.  
Very recently, Torres et al.~\cite{Torres:2014bpa} numerically simulated cosmological nonlinear structure formation. They did a comparative study of an oscillating massive scalar field and dust in spherical symmetry.

However, nonspherical effects on PBH formation should be significant if the effective equation of state is very soft, as indicated by Carr~\cite{Carr:1975qj}. This is 
because the Jeans scale is much smaller than the Hubble length and the deviation from spherical symmetry grows in Newtonian gravitational collapse of a uniform spheroid, which is known as the Lin--Mestel--Shu instability ~\cite{Lin:1965}. Nonspherical effects on PBH formation have been studied by Khlopov and Polnarev~\cite{Khlopov:1980mg} in the context of the phase transition of grand unification and their result has been applied to the ending phase of inflation~\cite{Alabidi:2012ex, Alabidi:2013lya}.

To study nonspherical effects on PBHs in a fully general relativistic,
analytical, and exact manner, we focus on dynamical exact solutions with
dust as an idealized model of a matter field with very soft effective equation of state. In the present context, we pay attention to the sequence of exact solutions, the Friedmann--Lemaitre--Robertson--Walker (FLRW) solution, the Lemaitre--Tolman--Bondi (LTB) solution~\cite{Lemaitre:1933, Tolman:1933, Bondi:1947}, and Szekeres's quasispherical solution~\cite{Szekeres:1974ct}. The FLRW solution is a spatially homogeneous and isotropic solution with no arbitrary function. It can also describe the interior of a uniform density dust ball~\cite{Oppenheimer:1939ue}. The LTB solution is a spherically symmetric and inhomogeneous solution with two arbitrary functions of one variable. This solution is particularly important because it is a general solution in spherical symmetry and includes the FLRW dust solution. The spherical model of PBH formation has been constructed with the LTB solution by Harada, Goymer, and Carr~\cite{Harada:2001kc}. The Szekeres solution is a nonspherically symmetric and inhomogeneous solution with five arbitrary functions of one variable. It includes the LTB solution and it admits no Killing vector in general. However, the whole set of Szekeres solutions is a proper subset of the whole set of general dust solutions, of which the explicit expression is not known. The Szekeres solution has been intensively studied in the context of nonlinear perturbation of the Universe~\cite{Ishak:2007rp,Nwankwo:2010mx,Ishak:2011hz,Peel:2012vg}.

A singularity which is not behind a black hole horizon, called a naked singularity, may appear in the course of gravitational collapse with regular initial data in the LTB solution~\cite{Yodzis:1973gha,Eardley:1978tr,Christodoulou:1984mz,Joshi:1993zg,Singh:1994tb,Jhingan:1996jb}. This is also the case in the Szekeres solution~\cite{Szekeres:1975dx,Joshi:1996qc,Deshingkar:1998ge,Nolan:2007gy}. 
Clearly this departure from spherical symmetry cannot avoid the formation of
naked singularities. Goncalves~\cite{Goncalves:2001yp} studied the cosmic censorship
and the curvature strength for the shell-focusing singularities in the 
quasispherical Szekeres solutions with dust and a cosmological constant
based on the radial null geodesics, which do not exist in general.
Hellaby and Krasinski~\cite{Hellaby:2002nx} studied the condition for the 
avoidance of shell-crossing singularities in the Szekeres solutions 
for the quasispherical, quasi-pseudospherical and quasiplanar cases
in the very general formulations. The same authors~\cite{Hellaby:2007hq} 
presented the geometrical
interpretation of the Szekeres solutions for the quasi-pseudospherical
and quasiplanar cases. 
Krasinski and Bolejko~\cite{Krasinski:2012hv} defined an 
absolute apparent horizon in the quasispherical
Szekeres solutions, discussed its difference from an apparent
horizon, which is commonly used, and concluded that the apparent
horizon can be regarded as the true horizon. 
Vrba and Svitek~\cite{Vrba:2014dwa} rewrote the
condition for the occurrence of shell-crossing singularities in terms of
the maximum, minimum, and average density of the shell at the moment of
occurrence. They also discussed the time evolution of the solutions, 
restricting themselves to the marginally bound case which is not 
relevant to the cosmological growing perturbation. 

It is important to see the condition for the
occurrence and non-occurrence of naked singularities and its physical
interpretation in the context of the nonlinear evolution of
cosmological primordial fluctuations. 
We adopt the simultaneous big bang condition to extract a growing
mode of adiabatic perturbations.
The formation and evolution of cosmological nonlinear perturbations 
have been analyzed on the same 
grounds~\cite{Tomita:1999rw,GarciaBellido:2008nz}.
For these nonlinear cosmological perturbations, 
we investigate the formation of black holes and cosmological voids
against the formation of shell-focusing and shell-crossing singularities
and see how the deviation from spherical symmetry affects these 
physical situations for the first time.
We also discuss the link between the current result and 
the physical interpretation by Khlopov and
Polnarev~\cite{Khlopov:1980mg} 
that shell-focusing naked singularities may be 
physically regarded as the onset of strong pressure gradient
force.

This paper is organized as follows. In Sect.~2, we present and interpret
the Szekeres solution. In Sect.~3, we describe the dynamics of the
Szekeres solution. In Sect.~4, we analyze the shell-crossing and
shell-focusing singularities under the simultaneous big bang
condition. There, we find a necessary and sufficient condition for the
Szekeres solution to describe the spherical and nonspherical formation
of PBHs without suffering naked singularity formation. In Sect.~5, we
propose several concrete models that describe the spherical and
nonspherical formation of PBHs without naked singularities. Section 6 is
devoted to our conclusions. In Appendix A, we briefly review the
derivation of the Szekeres solution in order for the paper to be self-contained. We use geometrized units, where $c=G=1$.

\section{Szekeres's quasispherical dust solution and its physical interpretation}

\subsection{Presentation of the metric}

We present here a brief overview of Szekeres's quasispherical dust solution. For completeness, the derivation of this solution is given in Appendix~\ref{sec:derivation_szekeres}. The line element is given by 
\begin{equation}
ds^{2}=-dt^{2}+e^{2\lambda}dr^{2}+e^{2\omega}(dx^{2}+dy^{2}),
\label{eq:metric}
\end{equation}
where 
\begin{eqnarray}
&& e^{\omega}=\frac{\phi}{P}, 
\label{eq:omega}\\
&& e^{\lambda}=P\frac{(e^{\omega})_{,1}}{W(r)}, 
\label{eq:lambda}\\
&& P=A(r)(x^{2}+y^{2})+2B_{1}(r)x+2B_{2}(r)y+C(r), 
\label{eq:P} \\
&& A(r)C(r)-B_{1}^{2}(r)-B_{2}^{2}(r)=\frac{1}{4}, \label{eq:algebraic_constraint}\\ 
&& (\phi_{,0})^{2}=W^{2}(r)-1+\frac{S(r)}{\phi}, 
\label{eq:Friedmann_spherical}\\
&& \rho=\frac{PS(r)'-3S(r)P_{,1}}{8\pi
 \phi^{2}[P\phi_{,1}-\phi P_{,1}]}, \label{eq:density} 
\end{eqnarray}
$(x^{0},x^{1},x^{2},x^{3})=(t,r,x,y)$, and $\phi=\phi(t,r)$.
The prime denotes the ordinary derivative with respect to the argument
and hence is only used for functions of one variable such as $A(r)$, $B_{1}(r)$, $B_{2}(r)$, $C(r)$, 
$W(r)$, and $S(r)$. For functions of more than one variable, the partial
derivatives are denoted by a comma followed by the index of the differentiating variable. Equation~(\ref{eq:Friedmann_spherical}) can be integrated to give
\begin{equation}
 \pm \int^{\phi}_{0}\displaystyle\frac{d \varphi}{\sqrt{W^{2}(r)-1+\frac{S(r)}{\varphi}}}=t+H(r),
\label{eq:H_QS}
\end{equation}
where $H(r)$ is an arbitrary function of $r$. Thus, the solution contains seven arbitrary functions of $r$. With one scaling freedom and one algebraic constraint taken into account, the solution contains five arbitrary functions. 

Equation~(\ref{eq:algebraic_constraint}) implies that neither $A(r)$ nor $C(r)$ can vanish, while $B_{1}(r)$ or $B_{2}(r)$ can. We assume $A(r)>0$ and $C(r)>0$ without loss of generality. Equation~(\ref{eq:P}) can be transformed 
into the following form:
\begin{equation}
 P=\frac{1}{4A(r)}\left[\left\{2A(r)\left(x+\frac{B_{1}(r)}{A(r)}\right)\right\}^{2}
+\left\{2A(r)\left(y+\frac{B_{2}(r)}{A(r)}\right)\right\}^{2}
+1 \right],
\label{eq:P_completing_the_square}
\end{equation}
which implies
$
 P\ge 1/[4A(r)]
$
and hence $P$ too is positive definite.

It is clear that the solution reduces to the LTB solution if $A(r)=C(r)=1/2$ and $B_{1}(r)=B_{2}(r)=0$. See Refs.~\cite{Landau:1982dva,Plebanski:2006sd} for the LTB solution, where $W^{2}(r)-1$, $S(r)$, and $H(r)$ correspond to the energy, mass, and time functions, respectively, and $\phi(t,r)$ gives the areal radius of the two-sphere of constant $t$ and $r$. The coordinates $(x,y)$ on the two-sphere can be understood in terms of the stereographic projection. One of the three arbitrary functions corresponds to the gauge freedom. For example, $H(r)$ can be fixed if we fix the radial coordinate at $t=t_{0}$ so that $\phi(t_{0},r)=r$. The simplest choice to recover the Minkowski spacetime is $S(r)=0$, $W(r)=1$, and $H(r)=0$.

The energy density can be singular only if 
\begin{equation}
\phi(t,r)=0
\end{equation}
or 
\begin{equation}
\phi(t,r)_{,1}P(r,x,y)-\phi(t,r)P(r,x,y)_{,1}=0 
\end{equation}
is satisfied. The former and the latter are called shell-focusing and shell-crossing singularities, respectively.

\begin{figure}[htbp]
 \begin{center}
  \includegraphics[width=0.5\textwidth]{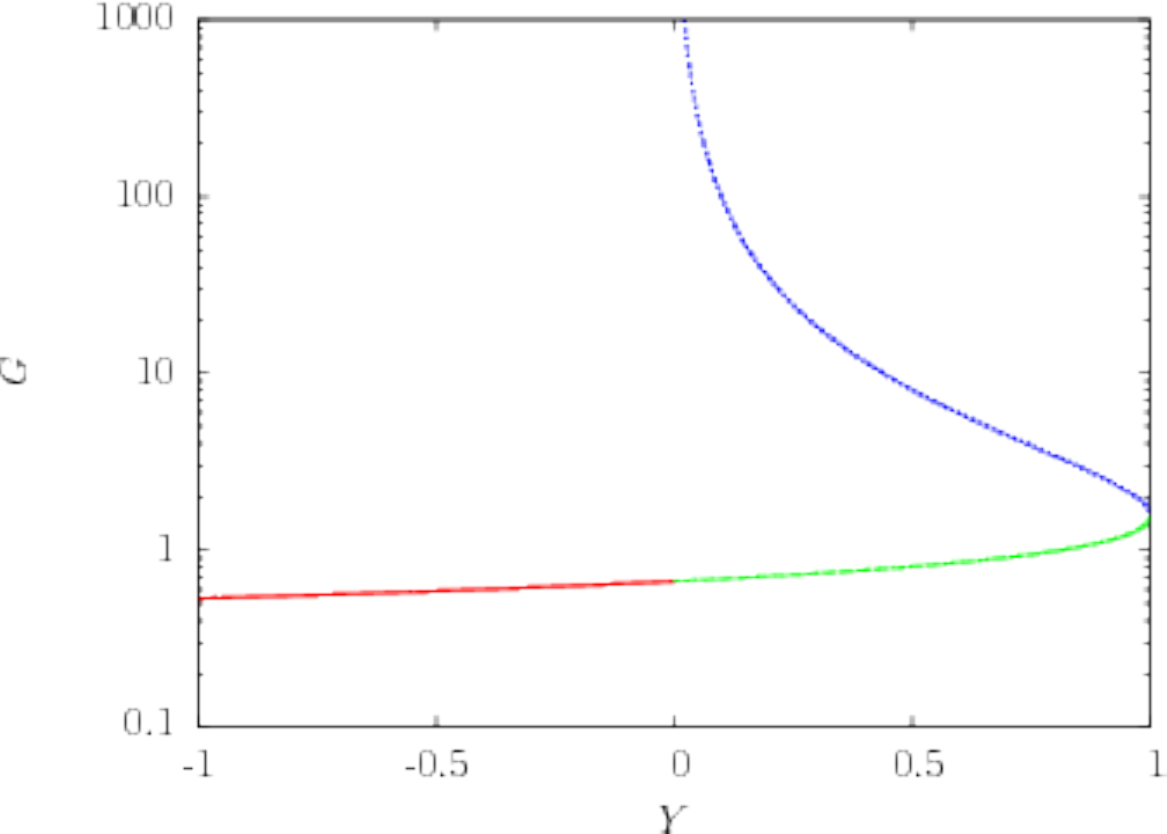}
\caption{\label{fg:G_generator}
The function $G$ is plotted, where only the branches which start with a big bang are chosen. The red and green lines denote $G_{+}$, while the blue line denotes $G_{-}$.}
 \end{center}
\end{figure}

Equation~(\ref{eq:H_QS}) describing the evolution of shells can be integrated explicitly. For $W(r)=1$, we have 
\begin{equation}
 t+H(r)=\pm \frac{2}{3}\sqrt{\frac{\phi^{3}}{S(r)}}.
\end{equation}
For $0<W(r)<1$, we have
\begin{eqnarray}
 \phi=\frac{S}{1-W^{2}(r)}\frac{1-\cos\eta}{2}, ~~
 t+H(r)=\frac{S(r)}{(1-W^{2}(r))^{3/2}}\frac{\eta-\sin\eta}{2}.
\end{eqnarray}
For $W(r)>1$, we have
\begin{eqnarray}
 \phi=\frac{S(r)}{W^{2}(r)-1}\frac{\cosh\eta-1}{2}, ~~
 t+H(r)=\frac{S(r)}{(W^{2}(r)-1)^{3/2}}\frac{\sinh\eta-\eta}{2}.
\end{eqnarray}
The above solutions can be summarized into the following compact form:
\begin{equation}
t+H(r)=\sqrt{\frac{\phi^{3}}{S(r)}}G\left(\frac{1-W^{2}(r)}{S(r)}\phi\right),
\label{eq:t+H=G} 
\end{equation}  
where 
\begin{eqnarray}
 G(Y)=\left\{\begin{array}{cc}
    G_{+}(Y)=\displaystyle\frac{\arcsin\sqrt{Y}}{Y^{3/2}}-\frac{\sqrt{1-Y}}{Y}& (0<Y\le 1,
     \mbox{expanding})\\
    G_{-}(Y)=\displaystyle\frac{\pi-\arcsin\sqrt{Y}}{Y^{3/2}}+\frac{\sqrt{1-Y}}{Y} & (0<Y\le
     1, \mbox{collapsing}) \\
    G_{+}(Y)=\frac{2}{3} & (Y=0,\mbox{expanding}) \\
    G_{-}(Y)=-\frac{2}{3} & (Y=0,\mbox{collapsing}) \\
    G_{+}(Y)=\displaystyle\frac{-\mbox{arcsinh}\sqrt{-Y}}{(-Y)^{3/2}}-\frac{\sqrt{1-Y}}{Y}
     & (Y<0,\mbox{expanding}) \\
    G_{-}(Y)=-\left[\displaystyle\frac{-\mbox{arcsinh}\sqrt{-Y}}{(-Y)^{3/2}}-\frac{\sqrt{1-Y}}{Y}\right] & (Y<0,\mbox{collapsing})
	  \end{array}
\right. ,
\label{eq:G_y}
\end{eqnarray}
and where
\begin{equation}
 Y:=\frac{1-W^{2}(r)}{S(r)}\phi.
\label{eq:Y}
\end{equation}
Note that for $0<W(r)<1$ the expanding and collapsing branches are combined into a single complete solution with both big bang and big crunch. Since we are interested in the cosmological solutions, we focus on the branches which possess a big bang, so that $G=G_{+}$ for $W(r)\ge 1$, while $G=G_{\pm}$ for $0<W(r)<1$. Figure~\ref{fg:G_generator} shows the relevant branches of $G$ for $-\infty<Y\le 1$. We note that $G_{-}(Y)$ for $0<W(r)<1$ admits the following expansion:
\begin{equation}
 G_{-}(Y)=\frac{\pi}{Y^{3/2}}-\frac{2}{3}-\frac{1}{5}Y-\frac{3}{28}Y^{2}+O(Y^{3}).
\label{eq:G_-_Y_exp}
\end{equation}

\subsection{Comoving coordinates and the dipole moment}

Inspired by the functional form of Eq.~(\ref{eq:P_completing_the_square}),
we make two successive coordinate transformations on the constant $t$ hypersurface. 
First from $(r,x,y)$ to new coordinates $(r,p,q)$, where $p$ and $q$ are defined by
\begin{equation}
 p:=2A(r)\left(x+\frac{B_{1}(r)}{A(r)}\right)
~~
\mbox{and} 
~~
 q:=2A(r)\left(y+\frac{B_{2}(r)}{A(r)}\right),
\end{equation}
and then from $(p, q)$ to $(\theta, \varphi)$ based on the stereographic projection, where
\begin{eqnarray}
 p=\cot\frac{\theta}{2}\cos\varphi ~~\mbox{and}~~
 q=\cot\frac{\theta}{2}\sin\varphi.
 \end{eqnarray}
The function $P$ can now be expressed as 
\begin{equation}
 P=\frac{1}{4A \sin^{2}\frac{\theta}{2}},
\label{eq:P_Lagrangian_coordinates}
\end{equation}
and the metric induced on the two-surface $S_{t,r}$ is given by the standard form of the two-sphere 
\begin{equation}
 ds^{2}|_{S_{t,r}}=\phi^{2}(t,r)(d\theta^{2}+\sin^{2}\theta d\varphi^{2}),
\end{equation}
whereas the line element in the spacetime contains off-diagonal terms  
$drd\theta$ and $drd\varphi$ in general.
Thus, we establish the following interesting relation:
\begin{equation}
 dM=\rho e^{\lambda+2\omega}drdxdy=\tilde{\rho}r^{2}\sin\theta
  dr d\theta d\varphi,
\label{eq:mass_conservation}
\end{equation}
where $M$ is the proper mass and $\tilde{\rho}$ is given by 
$
 \tilde{\rho}:=\rho e^{\lambda}{\phi^{2}}/{r^{2}}
$
and is calculated to give
\begin{equation}
 \tilde{\rho}=\frac{S'}{8\pi r^{2}W}\left(1-3\frac{S}{S'}\frac{P_{,1}}{P}\right).
\label{eq:tilde_rho_expression}
\end{equation}
Therefore, the mass contained within the volume element $dV=r^{2}\sin\theta
dr d\theta d\varphi$ is constant in time. This means the coordinates $(r, \theta, \phi)$ play the role of the comoving spherical coordinates and  $\tilde{\rho}$ can be interpreted as the conserved mass density.

We have seen that $\tilde{\rho}$ gives the conserved mass density in terms of the comoving coordinates $(r, \theta, \varphi)$ or $(\tilde{x},\tilde{y},\tilde{z})$. This suggests that with $\tilde{\rho}$ we can define a conserved dipole moment. In the comoving spherical coordinates, $P$ can be transformed to the standard form given in Eq.~(\ref{eq:P_Lagrangian_coordinates}).

Thus, defining vectors ${\vec \beta}$ and ${\vec n}$ as
\begin{equation}
 {\vec \beta}:=-r\left(2\left(B_{1}'-\frac{A'}{A}B_{1}\right),2\left(B_{2}'-\frac{A'}{A}B_{2}\right),\frac{A'}{A}\right),
\end{equation}
and 
\begin{equation}
 {\vec n}:=\left(\sin\theta\cos\varphi,\sin\theta\sin\varphi,\cos\theta\right)=\frac{1}{r}(\tilde{x},\tilde{y},\tilde{z}),
\end{equation}
respectively, 
we find 
\begin{equation}
 \frac{P_{,1}}{P}=-{\vec n} \cdot \frac{{\vec \beta}}{r}
\label{eq:P_1_nbeta}
\end{equation}
in the coordinate basis of the comoving Cartesian coordinates $(\tilde{x},\tilde{y},\tilde{z})$. Since the nonspherical dependence of the 
density distribution $\tilde{\rho}$, as well as $\rho$, appears only
through this combination, the matter distribution has monopole and
dipole moments only and the vector ${\vec \beta}$ is proportional to the
dipole moment. The absolute value $\beta(r)$ of the vector ${\vec \beta}(r)$ is given by 
\begin{equation}
 \beta^{2}(r)=4r^{2}[B_{1}'^{2}(r)+B_{2}'^{2}(r)-A'(r)C'(r)].
\end{equation}
It can be easily shown that $\beta^{2}(r)$ is positive definite. The definition of $\beta$ used here is the same as in Szekeres~\cite{Szekeres:1975dx}.

We can rewrite the expression for the energy density, Eq.~(\ref{eq:tilde_rho_expression}), in the following form:
\begin{equation}
 \tilde{\rho}=\tilde{\rho}_{0}\left(1+{\vec n}\cdot {\vec d}\right),
\end{equation}
where 
\begin{eqnarray}
 \tilde{\rho}_{0}:=\frac{S'}{8\pi r^{2}W} ~~\mbox{and}~~
 {\vec d}:=3\frac{S}{rS'}{\vec \beta}.
\end{eqnarray}
Thus, we can naturally define the nondimensional mass dipole moment localized in the shell labeled $r$ with ${\vec d}$.

On the other hand, the physical density $\rho$ can also be written in the following form:
\begin{equation}
 \rho=\rho_{0}\left(1+\chi \right),
\end{equation}
where 
\begin{eqnarray}
 \rho_{0}:=\frac{S'}{8\pi \phi^{2}\phi_{,1}}~~\mbox{and}~~
 \chi:
=\frac{\left(\frac{S'}{S}-3\frac{\phi_{,1}}{\phi}\right)\frac{P_{,1}}{P}}
{\left(\frac{\phi_{,1}}{\phi}-\frac{P_{,1}}{P}\right)\frac{S'}{S}}.
\label{eq:chi}
\end{eqnarray}
We can interpret $\rho_{0}$ as the spherical part of the density field, which is identical to that of the reference LTB solution, and $\chi$ as the deviation from it. The nonsphericity $\chi$ does depend on time through $\phi_{,1}/\phi$. Although $\chi$ cannot be simply interpreted as the deviation due to the dipole moment, it is closely related to the vector ${\vec \beta}$. If ${\vec \beta}={\vec 0}$, then $\chi$ vanishes identically. Conversely, if $\chi$ vanishes identically, then ${\vec \beta}=0$ or $\phi_{,1}/\phi=S'/(3S)$.

\subsection{Spherically symmetric and axially symmetric subclasses}
As is seen in Eq.~(\ref{eq:density}), the energy density depends only on $t$ and $r$, if 
\begin{equation}
 \frac{A'}{A}=\frac{B_{1}'}{B_{1}}=\frac{B_{2}'}{B_{2}}=\frac{C'}{C}
\label{eq:spherical_case}
\end{equation}
or
\begin{equation}
 \frac{S'}{3S}=\frac{\phi_{,1}}{\phi}
\label{eq:separable_case}
\end{equation}
is satisfied~\cite{Bonnor:1976zz}. From Eqs.~(\ref{eq:algebraic_constraint}), 
Eq.~(\ref{eq:spherical_case}) implies that all of $A(r)$, $B_{1}(r)$, $B_{2}(r)$, and $C(r)$ are constant. Then, the spacetime becomes spherically symmetric and the solution reduces to the LTB solution. In this case, we can see ${\vec \beta}={\vec 0}$. This supports our interpretation of ${\vec \beta}$ as a dipole moment. Equation~(\ref{eq:separable_case}) holds if and only if $\phi$ is separable as we can show from Eq.~(\ref{eq:Friedmann_spherical}). In this case, $\phi=a(t)S^{2/3}(r)$ and therefore $\rho=3/[8\pi a^{3}(t)]$, i.e., the density is homogeneous. In this case, since the spatial component of the metric is written as $g_{ij}=a(t)\gamma_{ij}(x^{i})$, $\gamma_{ij}$ can be shown to be  that of the constant curvature space in three dimensions and therefore this is the FLRW solution~\cite{Szekeres:1974ct}.

The above discussion also implies that the Szekeres solution can be matched to the Schwarzschild solution at any radius $r=r_{m}$. In fact, we can always choose $W(r)$, $S(r)$, $H(r)$, $A(r)$, $B_{1}(r)$, $B_{2}(r)$, $C(r)$ to be constant for $r> r_{m}$.
The mass parameter of the Schwarzschild black hole is given by $M=S(r_{m})/2$. Then, the region for $r> r_{m}$ is spherically symmetric vacuum and hence the Schwarzschild solution by Birkhoff's theorem.

Next, let us consider a subclass where $B_{1}(r)=B_{2}(r)=0$ identically but $A(r)$ is still allowed to be a function of $r$. Then, Eqs.~(\ref{eq:P}) and (\ref{eq:algebraic_constraint}) reduce to 
\begin{equation}
 P(r,x,y)=A(r)(x^{2}+y^{2})+\frac{1}{4A(r)}.
\end{equation}
In the comoving spherical coordinates $(r,\theta,\varphi)$, we can easily find 
\begin{equation}
 \frac{P_{,1}}{P}=\frac{A'}{A}\cos\theta.
\end{equation}
Since the metric components in the comoving spherical coordinates do not depend on $\varphi$, the spacetime is axially symmetric. In the comoving Cartesian coordinates $(\tilde{x},\tilde{y},\tilde{z})$, we find 
\begin{equation}
 {\vec \beta}=\left(0,0,-r\frac{A'}{A}\right)
\end{equation}
Thus, ${\vec \beta}$ is directed along the $\tilde{z}$ axis. This also supports our interpretation of ${\vec \beta}$ as a dipole moment. 

\section{Dynamics of the Szekeres solution}

\subsection{Shell-crossing singularities}

Following Szekeres~\cite{Szekeres:1975dx}, 
we quote two lemmas for quadratic forms.
\begin{Lem}\label{lemma:1}
A quadratic form $Q=a \xi\bar{\xi}+b\xi+\bar{b}\bar{\xi}+c$ ($a\ne 0$), where $a$ and $c$ are real and $b$ is complex, has no zeros in the complex plane $\xi$ if and only if its discriminant is positive, i.e., $\Delta(Q)=ac-b\bar{b}>0$. Moreover, if and only if $a>0$ and $\Delta(Q)>0$, $Q>0$ for any $\xi$. 
\end{Lem}
\begin{Lem}\label{lemma:2}
\begin{equation}
 \Delta(\theta'P-c\theta P_{,1})=\frac{1}{4}\theta'^{2}+c^{2}\theta^{2}\Delta(P_{,1}),
\end{equation}
where $c$ is a constant and $\theta=\theta(r)$. 
\end{Lem}

If we denote the time of the shell-crossing singularity by $t=t_{\rm SC}(r,x,y)$, where 
\begin{equation}
P(r,x,y)\phi_{,1}(t_{\rm SC}(r,x,y),r)-\phi(t_{\rm SC}(r,x,y),r) P_{,1}(r,x,y)=0,
\label{eq:shell-crossing} 
\end{equation}
$t_{\rm SC}$ depends not only on $r$ but also on $x$ and $y$ in general. Thus, shell-crossing singularities are affected by nonsphericity. If we fix $t$, a shell-crossing singularity occurs not on the two-surface $S_{t,r}$ but on a different two-surface in general. From Eq.~(\ref{eq:chi}), we can find that if $P_{,1}\ne 0$ and $\phi_{,1}/\phi\ne S'/(3S)$, then $\chi$ diverges to infinity at shell-crossing singularities. In other words, if the shell is nonspherical, it is the nonspherical rather than the spherical part of the density field that diverges at the shell-crossing
singularities. Therefore, generic shell-crossing singularities in the Szekeres 
spacetime are essentially nonspherical. 

Applying the lemmas presented above  to the function $\phi_{,1}P-\phi P_{,1}$, we can show that the two-surface $S_{t,r}$ possesses a shell-crossing singularity if and only if 
\begin{equation}
\left(\frac{\phi_{,1}}{\phi}\right)^{2}\le
 4[(B_{1}')^{2}+(B_{2}')^{2}-A'C'] ,
\label{eq:shell-crossing_occurrence}
\end{equation}
which can be rewritten as
\begin{equation}
\left(\frac{\phi_{,1}}{\phi}\right)^{2}\le \frac{\beta^{2}}{r^{2}} ,
\label{eq:shell-crossing_occurrence_beta}
\end{equation}
directly linking the shell-crossing condition with nonsphericity parameter
$\beta$.
We define $t_{\rm SC}(r)$ as the time of the {\it earliest} occurrence of
shell-crossing singularity on $S_{t,r}$ so that the quadratic form
$P\phi_{,1}-\phi P_{,1}$ with fixed $r$ 
begins to have a zero at $t=t_{\rm SC}(r)$. This implies 
\begin{equation}
\left.\left(\frac{\phi_{,1}}{\phi}\right)^{2}\right|_{t=t_{\rm SC}(r)}
= \frac{\beta^{2}(r)}{r^{2}}.
\label{eq:shell-crossing_first_occurrence_beta}
\end{equation}

\subsection{Regularity condition}

We assume the existence of a regular initial data surface ($t=t_{0}$). Since the areal radius of the two-surface $S_{t,r}$ is $\phi(t,r)$, we also assume that $\phi_{0}(r):=\phi(t_{0},r)$ is an increasing function of $r$ and scale $r$ so that 
$\phi_{0}(r)=O(r)$. For the center to be locally Minkowski, Eq.~(\ref{eq:omega})
implies $W(0)=1$. For $\phi_{,0}$ to be bounded, Eq.~(\ref{eq:Friedmann_spherical}) implies $S(r)=O(r)$. Then, for $\rho_{0}(r):=\rho(t_{0},r)$ to be bounded as $r\to 0$, Eq.~(\ref{eq:density}) implies $S'(r)=O(r^{2})$. This implies 
\begin{equation}
S(r)=O(r^{3})\quad \mbox{and}\quad  1-W^{2}(r)=O(r^{2}).
\label{eq:regularity_1}
\end{equation}

We further assume that the metric is $C^{2-}$ in the comoving Cartesian coordinates. For $rP_{,1}/P$ to be $C^{2-}$ in terms of $(\tilde{x},\tilde{y},\tilde{z})$, $\beta(r)=O(r)$, and hence the nonspherical functions $A(r)$, $B_{1}(r)$, $B_{2}(r)$, and $C(r)$ are continuous and differentiable at $r=0$.

At $t=t_{0}$, the initial data is shell-crossing free for all $r(>0)$, $x$, and $y$ if, and only if 
\begin{equation}
 \left(\frac{\phi_{0,1}}{\phi_{0}}\right)^{2}>\frac{\beta^{2}(r)}{r^{2}}.
\label{eq:no_shell-crossing_general}
\end{equation}
In turn, applying Lemma~\ref{lemma:2} to the function $S'P-3SP_{,1}$, we can see that $\rho$ is positive definite at $t=t_{0}$ if, and only if 
\begin{equation}
 \frac{\beta^{2}(r)}{4r^{2}}<\frac{(S')^{2}}{36S^{2}}.
\end{equation}
Thus, regularity imposes the following condition on $\beta$:
\begin{equation}
 \beta(r)<\min \left(\frac{r\phi_{0,1}(r)}{\phi_{0}(r)}, r(\ln
	    S^{1/3}(r))'\right). 
\label{eq:regularity_3}
\end{equation}
The above condition is automatically satisfied near $r=0$, since $\beta(r)=O(r)$.

\subsection{Trapped surfaces, and apparent horizons}

Following Szekeres~\cite{Szekeres:1975dx}, we consider a trapping condition for a 
spacelike two-surface $S_{t,r}$. Since the tangent space of $S_{t,r}$ is spanned by $\partial/\partial x$, and $\partial/\partial y$, any normal vector $n^{\mu}$ to $S_{t,r}$ should satisfy $n^{2}=n^{3}=0$. Thus, if we consider the congruence of null geodesics with tangent vector $k^{\mu}$ normal to $S_{t,r}$, we find 
\begin{eqnarray}
 k^{\mu}k_{\mu}&=&0, \label{eq:null}\\
 k^{\mu}_{~;\nu}k^{\nu}&=& 0 \label{eq:geodesic}
\end{eqnarray}
with  $k^{2}=k^{3}=0$ on $S_{t,r}$. Assuming $k^{0}>0$, we can identify the null geodesics of $k^{1}>0$ ($<0$) with outgoing (ingoing) ones. We can choose $k^{1}=1$ ($-1$) on $S_{t,r}$ by choosing the scaling of the affine parameter. The sign of the expansion coincides with the sign of $k^{\mu}_{~;\mu}$, which is calculated to give
\begin{equation}
 k^{\mu}_{~;\mu}=k^{0}_{~,0}+k^{1}_{~,1}+\left(\lambda_{,0}+2\omega_{,0}\right)e^{\lambda}+\left(\lambda_{,1}+2\omega_{,1}\right)\epsilon,
\label{eq:expansion}
\end{equation}
where we put $k^{1}=\epsilon$,, and $\epsilon=\pm 1$. Differentiating  Eq.~(\ref{eq:null}) with respect to $t$, we find 
\begin{equation}
 k^{0}_{~,0}-\epsilon
  e^{\lambda}k^{1}_{~,0}-e^{\lambda}\lambda_{~,0}=0.
\label{eq:null'}
\end{equation}
In Eq.~(\ref{eq:geodesic}) with $\mu=1$, we find 
\begin{equation}
 e^{\lambda}k^{1}_{~,0}+\epsilon(k^{1}_{~,1}+2\lambda_{~,0}
e^{\lambda})+\lambda_{,1}=0.
\label{eq:geodesic'}
\end{equation}
We eliminate $k^{1}_{~,0}$ from Eq.~(\ref{eq:null'}) by Eq.~(\ref{eq:geodesic'}),, and eliminate $k^{0}_{~,0}+k^{1}_{~,1}$ from Eq.~(\ref{eq:expansion}) by the resultant equation.  Then, we find 
\begin{equation}
 k^{\mu}_{~;\mu}=2\omega_{,0}e^{\lambda}+2\epsilon \omega_{,1}.
\label{eq:null_expansion}
\end{equation}
Using Eqs.~(\ref{eq:omega}), (\ref{eq:lambda}), and (\ref{eq:Friedmann_spherical}), we can transform Eq.~(\ref{eq:null_expansion}) to
\begin{equation}
 k^{\mu}_{~;\mu}=2\left(\frac{\phi_{,1}}{\phi}-\frac{P_{,1}}{P}\right)
\left(
\iota \sqrt{1+\frac{1}{W^{2}}\left(\frac{S}{\phi}-1\right)}
+\epsilon\right),
\end{equation}
where $\iota=\mbox{sign}
(\phi_{,0})$. Equation~(\ref{eq:no_shell-crossing_general}) implies that
the first factor is positive because it cannot change the sign without
encountering a shell-crossing singularity. Thus, we establish that if the dust is collapsing ($\iota=-1$), the outgoing ($\epsilon=1$) null normal can have vanishing expansion if $S=\phi$, while if the dust is expanding ($\iota=1$), it is the ingoing ($\epsilon=-1$) null normal that can have vanishing expansion if $S=\phi$.

Here we identify a marginally trapped two-surface $S_{t,r}$ with an
apparent horizon according to Krasinski and
Bolejko~\cite{Krasinski:2012hv}. Then we call an apparent horizon with
vanishing outgoing (ingoing) null expansion a future (past) apparent
horizon. Note that this is somewhat different from the notion of an
apparent horizon defined by Hawking and
Ellis~\cite{Hawking:1973uf}. Although a black hole horizon is usually
identified with an event horizon in the asymptotically flat spacetime,
such an identification is not so strongly motivated in the cosmological
spacetime because of the teleological nature of an event horizon and
because the asymptotic condition at null infinity is less physically
meaningful in cosmology with a finite particle horizon. 
Although the local and dynamical definition of black hole horizon is ambiguous, we can identify a future apparent horizon with a local black hole horizon in this paper for not only its physical reasonableness but also its simplicity and usefulness for the analysis. 

In the dust models, a noncentral shell-focusing singularity is always future trapped because $\phi(t,r)=0$ and $S(r)>0$ there. This means that noncentral singularity is causally disconnected from a distant observer.

\subsection{Six critical events}

As for the dynamics of the Szekeres solution, there are six important events: big bang, past apparent horizon, maximum expansion, future apparent horizon, big crunch and shell-crossing singularity. We denote the times of the occurrence of these six events at each shell with $t_{\rm BB}(r)$, $t_{\rm PH}(r)$, $t_{\rm ME}(r)$, $t_{\rm FH}(r)$, $t_{\rm BC}(r)$, and 
$t_{\rm SC}(r)$, respectively.

The order of these events is trivial except for shell-crossing
singularity. This is because, except for shell-crossing singularity, the events are determined solely by each shell and its dynamics is identical to the FLRW spacetime.

For $W(r)\ge 1$, we can find $t_{\rm BB}(r)<t_{\rm PH}(r)$ and there is no maximum expansion, no future apparent horizon and no big crunch. For the big bang time, Eq.~(\ref{eq:t+H=G}) implies 
\begin{equation}
t_{\rm BB}(r)=-H(r).
\label{eq:tBB}
\end{equation}
Since $\phi=S$ at apparent horizons, 
we find
\begin{equation}
 t_{\rm PH}(r)-t_{\rm BB}(r)=S(r)G_{+}(1-W^{2}(r)).
\label{eq:tPH}
\end{equation}

For $0<W(r)< 1$, we can find $t_{\rm BB}(r)<t_{\rm PH}(r)<t_{\rm ME}(r)<t_{\rm FH}(r)<t_{\rm BC}(r)$. The big bang time and the time of past apparent horizon are given by Eqs.~(\ref{eq:tBB}) and (\ref{eq:tPH}). The time of future apparent horizon is given by 
\begin{equation}
 t_{\rm FH}(r)-t_{\rm BB}(r)=S(r)G_{-}(1-W^{2}(r)).
\label{eq:t_FH}
\end{equation}
For the big crunch time, we find 
\begin{equation}
 t_{\rm BC}(r)-t_{\rm BB}(r)=\lim_{\phi\to 0}\sqrt{\frac{\phi^{3}}{S(r)}}G_{-}\left(\frac{1-W^{2}(r)}{S(r)}\phi\right),
\end{equation}
and the result is 
\begin{equation}
 t_{\rm BC}(r)-t_{\rm BB}(r)=\pi\frac{S(r)}{[1-W^{2}(r)]^{3/2}}.
\label{eq:t_BC}
\end{equation}
The time of maximum expansion is given by $\phi=S(r)/[1-W^{2}(r)]$, as seen from
Eq.~(\ref{eq:Friedmann_spherical}). Therefore, from Eqs.~(\ref{eq:t+H=G}) and (\ref{eq:G_y}), we find
\begin{equation}
 t_{\rm ME}(r)-t_{\rm BB}(r)=\frac{\pi}{2}\frac{S(r)}{[1-W^{2}(r)]^{3/2}}.
\end{equation}

The time of shell-crossing singularity is highly nontrivial because $t_{\rm SC}(r)$ is determined not solely by the dynamics of the shell labeled $r$, but that of the neighboring shells. We will study it in detail in the next section.

\section{Szekeres solution as nonlinear cosmological perturbations}

\subsection{Simultaneous big bang condition}

The simultaneous big bang condition is often used in the cosmological context and suitable for a nonlinear growing mode of adiabatic perturbation. This condition implies $H(r)=-t_{\rm BB}(r)=$ const. in Eq.~(\ref{eq:tBB}). If we assume this condition, we find from Eqs.~(\ref{eq:t+H=G}) and (\ref{eq:G_y})
\begin{eqnarray}
\phi\to \left(\frac{9S}{4}\right)^{1/3}(t-t_{\rm BB})^{2/3}~~\mbox{and}~~
 \rho\to \frac{1}{6\pi (t-t_{\rm BB})^{2}}
\end{eqnarray} 
in the limit of $t\to t_{\rm BB}$, and hence the solution approaches the flat FLRW solution.
This implies that the simultaneous big bang condition corresponds to
extracting a purely growing mode of adiabatic perturbation from the flat
FLRW universe. This condition in the LTB solution has been adopted for
the construction of the primordial black hole  formation model in the flat FLRW universe in Harada et al.~\cite{Harada:2001kc}. The LTB solution as an exact model of black holes in the evolving Universe is also discussed from a very broad scope in Sect. 18.9 of Ref.~\cite{Plebanski:2006sd}.

\subsection{Shell-crossing singularities}

We should note that very general treatments of
the occurrence of shell-crossing singularities 
for the LTB solution and the Szekeres solution
are given in Sects. 18.10 and 19.7.4 of Ref.~\cite{Plebanski:2006sd}, 
respectively. Our aim in the next few subsections is to 
analyze the occurrence of shell-crossing singularities 
and get a physical insight into the nature of 
shell-crossing occurrence in the present cosmological setting.

\subsubsection{Expression for $\psi=r\phi_{,1}/\phi$}

In order to see the occurrence of shell-crossing singularities, 
it is important to have the explicit form of $r\phi_{,1}/\phi$. 
In the general case, this can be found by differentiating
Eq.~(\ref{eq:t+H=G}) with respect to $r$ as follows:
\begin{equation}
\psi:= \frac{r\phi_{,1}}{\phi}=\frac{\displaystyle\frac{1}{S}\left(\frac{1-W^{2}}{Y}\right)^{3/2}rH'+\frac{1}{2}r(\ln S)'G-r\left(\ln \frac{|1-W^{2}|}{S}\right)'\displaystyle\frac{dG}{d\ln |Y|}}
{\displaystyle\frac{3}{2}G+\displaystyle\frac{dG}{d\ln |Y|}},
\label{eq:psi_general}
\end{equation}
where the derivative of $G$ is given by the following form:
\begin{eqnarray}
 \frac{dG}{dY}=\left\{\begin{array}{cc}
    \displaystyle\frac{dG_{+}}{dY}=-\frac{3}{2}\displaystyle\frac{\arcsin\sqrt{Y}}{Y^{5/2}}+\frac{3-Y}{2Y^{2}\sqrt{1-Y}}& (0<Y\le 1,
     \mbox{expanding})\\
    \displaystyle\frac{dG_{-}}{dY}=-\frac{3}{2}\displaystyle\frac{\pi-\arcsin\sqrt{Y}}{Y^{5/2}}-\frac{3-Y}{2Y^{2}\sqrt{1-Y}} & (0<Y\le
     1, \mbox{collapsing}) \\
    \displaystyle\frac{dG_{+}}{dY}=\displaystyle\frac{1}{5} & (Y=0, \mbox{expanding}) \\
    \displaystyle\frac{dG_{+}}{dY}=-\frac{3}{2}\displaystyle\frac{\mbox{arcsinh}\sqrt{-Y}}{(-Y)^{5/2}}+\frac{3-Y}{2Y^{2}\sqrt{1-Y}} & (Y<0,\mbox{expanding})
	  \end{array}
\right. .
\end{eqnarray}

\begin{figure}[htbp]
\begin{center}
\begin{tabular}{cc}
 \includegraphics[width=0.45\textwidth]{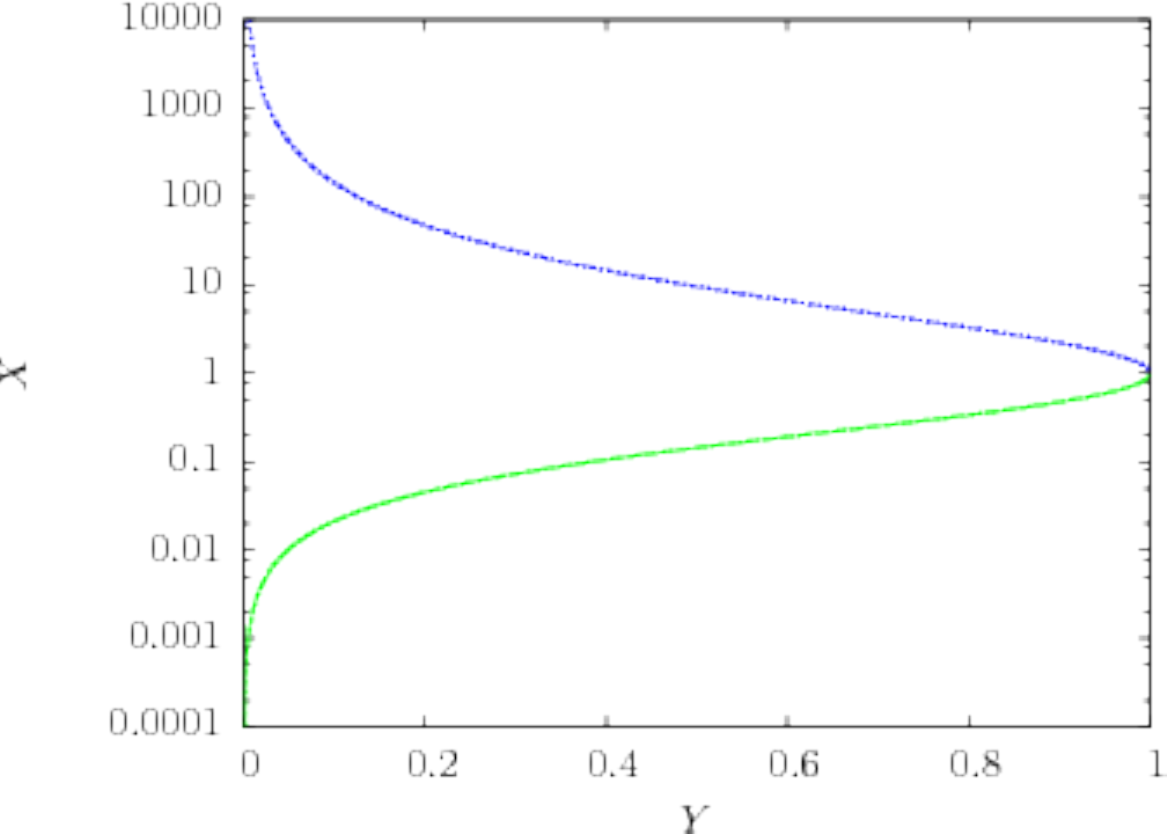}&
 \includegraphics[width=0.45\textwidth]{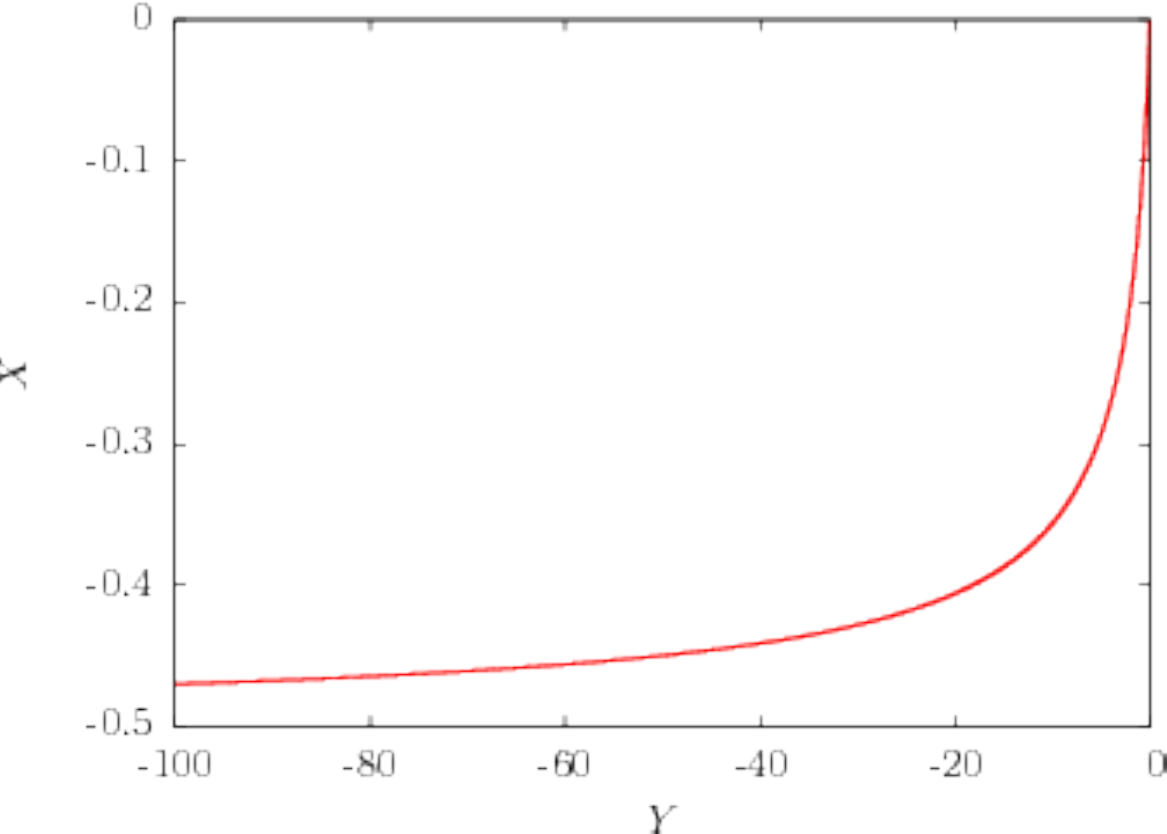} \\
\end{tabular}
\caption{\label{fig:X} The behavior of $X$ as a function of $Y$. The
 left and right panels show $X$ for $Y\ge 0$ and $Y\le 0$, respectively.
The red and green lines show $X_{+}$, while the blue line shows $X_{-}$.}
\end{center}
\end{figure}

With the 
simultaneous big bang condition $H(r)=-t_{\rm BB}=$const., 
Eq.~(\ref{eq:psi_general}) is reduced to 
\begin{equation}
 \psi=\frac{\displaystyle\frac{1}{2}r(\ln S)'-r\left(\ln
  \frac{|1-W^{2}|}{S}\right)'\displaystyle\frac{d\ln G}{d\ln |Y|}
}{\displaystyle\frac{3}{2}+\displaystyle\frac{d\ln
G}{d\ln |Y|}}.
\label{eq:psi}
\end{equation}
Putting 
\begin{eqnarray}
 X(Y)=X_{\pm}(Y):=\frac{2\displaystyle\frac{d\ln G_{\pm}}{d\ln |Y|}}{2\displaystyle\frac{d\ln G_{\pm}}{d\ln |Y|}+3}, ~~
 a_{1}(r):=-r\left[\ln\frac{|1-W^{2}|}{S^{2/3}}\right]',~~
 a_{0}(r):=r(\ln S^{1/3})',
\label{eq:a_0}
\end{eqnarray}
we can rewrite $\psi$ in the following form:
\begin{equation}
 \psi(r,Y)=a_{1}(r)X(Y)+a_{0}(r).
\label{eq:phi'/phi}
\end{equation}
The function $X(Y)$ is plotted in Fig.~\ref{fig:X}.

For $W(r)=1$, we can see that 
the expression (\ref{eq:phi'/phi}) is very singular. 
For this case, regularity is 
obvious in the following expression for $\psi$:
\begin{equation}
 \psi=2\frac{\displaystyle\frac{d\ln
  G}{dY}\left[-r(1-W^{2})'\displaystyle\frac{\phi}{S}+\displaystyle\frac{2}{3}Yr(\ln S)'\right]}
  {2Y\displaystyle\frac{d\ln G}{dY}+3}+a_{0}(r).
\label{eq:phi'/phi_regular}
\end{equation}
In the limit $W\to 1$ where $0<S<\infty$ and $0<\phi<\infty$ are fixed, 
we find $Y\to 0$ and $d\ln G/dY\to 3/10$. Therefore, for $W(r)=1$, we find 
\begin{equation}
 \psi(r,\tilde{X})=\tilde{a}_{1}(r)\tilde{X} +a_{0}(r),
\label{eq:phi'/phi_W=1}
\end{equation}
where 
\begin{eqnarray}
 \tilde{X}:=\frac{\phi}{S},~~ \tilde{a}_{1}(r):=-\frac{1}{5}r(1-W^{2})',
\end{eqnarray}
and $a_{0}(r)$ is the same as in Eq.~(\ref{eq:a_0}).

From Eq.~(\ref{eq:shell-crossing_occurrence_beta}), the two-surface $S_{t,r}$
possesses a shell-crossing if and only if
$
 \psi \le \beta(r).
$
According to Eq.~(\ref{eq:phi'/phi}) for $W\ne 1$ and 
Eq.~(\ref{eq:phi'/phi_W=1}) for $W=1$, we can now 
analyze rigorously the occurrence of shell-crossing singularity 
for given $t$ and $r$ under the simultaneous big bang condition.

We should note that 
Eq.~(\ref{eq:regularity_1}) implies
$
a_{1}=O(r^{2})
$
for $W\ne 1$ and 
$
\tilde{a}_{1}=O(r^{2})
$
for $W=1$.  
As for $a_{0}(r)$, Eq.~(\ref{eq:regularity_1}) implies
$a_{0}(r)\to 1$ as $r\to 0$.
Since $\beta(r)=O(r)$ implies
$
 a_{0}(r)>\beta(r)
$
for sufficiently small $r$,
there is no shell-crossing singularity for sufficiently small $r$. 
From Eqs.~(\ref{eq:phi'/phi}) and 
(\ref{eq:phi'/phi_W=1}), since $X$ for $W\ne 1$ and $\tilde{X}$ for $W=1$
both begin with 0 at the big bang, 
no shell-crossing singularity appears 
at least for a sufficiently short time interval after the big bang.

We now discuss in general the cases in terms of the shell labeled $r$ 
with $0<W(r)<1$, $W(r)>1$, and $W(r)=1$, separately.
Hereafter, 
for simplicity, we choose the scaling of $r$ so that $S(r)=S_{3}r^{3}$
in accordance with Eq.~(\ref{eq:regularity_1}), where
$S_{3}$ is a positive constant. The result does not
depend on the choice of the scaling.
Recall that Eq.~(\ref{eq:regularity_3}) holds for $\beta(r)$.

\subsubsection{Bound shell: $0<W(r)<1$}
For $0<W(r)<1$, as  
the shell labeled $r$ begins with a big bang, reaches maximum
expansion, and ends in a big crunch,
$Y$ monotonically increases from 0, takes value 1 at maximum expansion, 
and then monotonically decreases to 0.
Meanwhile, 
$X=X_{+}$ monotonically increases from $0$ to $1$, 
which is its value at maximum expansion, and then switches to the 
$X=X_{-}$ branch, which 
monotonically increases from $1$ to $\infty$ as time proceeds.
We should note that $X_{-}$ admits the following expansion:
\begin{equation}
 X_{-}(Y)=\frac{3\pi}{2}Y^{-3/2}-\frac{3\pi}{4}Y^{-1/2}
-\frac{3\pi}{16}Y^{1/2}+\frac{1}{5}Y+O(Y^{3/2}).
\label{eq:expansion_shell-focusing}
\end{equation}
So, if $a_{1}(r)>0$, $\psi$ monotonically increases from $a_{0}(r)$
to $\infty$ as time proceeds. Therefore, 
we can conclude that there appears no shell-crossing singularity. 
This is also the case if $a_{1}(r)=0$, where $\psi$
is constant in time. If $a_{1}(r)<0$, $\psi$ monotonically decreases
from $a_{0}(r)$ to $-\infty$. Since $\beta (r)\ge 0$, this means that 
a shell-crossing singularity necessarily appears before the big crunch.
In this case, it is important whether or not the shell-crossing
singularity appears before
the maximum expansion and the future apparent horizon. 
A maximum expansion is characterized by $Y=1$ or $X=1$.
Since $\psi$ is a monotonically decreasing function of time, we can
conclude that a shell-crossing singularity at $r$ appears before
or coinciding with a maximum expansion if and only if 
$a_{1}(r)+a_{0}(r)\le \beta(r)$.
A future apparent horizon 
is characterized by $\phi=S$ in the collapsing branch.
The value of $\psi$ on the future apparent horizon is given by 
$ \psi=a_{1}(r)X_{-}(1-W^{2}(r))+a_{0}(r)$.
Since $\psi$ is a monotonically decreasing function of time, we can
conclude that a shell-crossing singularity at $r$ appears before
or coinciding with a future apparent horizon if and only if  
\begin{equation}
a_{1}(r)X_{-}(1-W^{2}(r))+a_{0}(r)\le \beta(r).
\label{eq:shell_crossing_before_future_apparent_horizon}
\end{equation}
If $|1-W^{2}|/S^{2/3}$ is a monotonically decreasing function of $r$, 
$a_{1}(r)>0$ and hence there is no shell-crossing 
singularity. Otherwise, there exists $r(>0)$ for which 
$a_{1}(r)<0$ and a shell-crossing singularity appears.
Then,  
Eq.~(\ref{eq:shell_crossing_before_future_apparent_horizon})
determines whether or not the shell-crossing singularity appears before
or coinciding with a future apparent horizon.
We can see that a large dipole moment can promote and advance the occurrence of 
shell-crossing singularity before the future apparent horizon.

\subsubsection{Unbound shell: $W(r)> 1$}
If $W(r)>1$, the shell labeled $r$ begins with a big bang and expands forever. 
In this case, 
$Y$ monotonically decreases from $0$ to $-\infty$ 
and $X$ monotonically decreases from $0$ to $-1/2$.
So, if $a_{1}(r)<0$, $\psi$ monotonically increases from $a_{0}(r)$
to 
$-a_{1}(r)/2+a_{0}(r)$
as time proceeds. Therefore, 
we can conclude that no shell-crossing singularity appears.
This is also the case if $a_{1}(r)=0$, where $\psi$
is constant in time. If $a_{1}(r)>0$, $\psi$ monotonically decreases
from $a_{0}(r)$ to $-a_{1}(r)/2+a_{0}(r)$. 
Since $\beta(r)\ge 0$, this means that 
a shell-crossing singularity eventually appears in the course of
expansion if and only if $-a_{1}(r)/2+a_{0}(r)\le \beta(r)$, i.e., 
\begin{equation}
 \frac{1}{2}r\left[\ln |1-W^{2}|\right]' \le \beta(r).
\label{eq:shell-crossing_occurrence_W>1}
\end{equation}
This means that if $W$ has an extremum which is greater than 1, 
there necessarily appears a shell-crossing singularity, 
irrespective of the value of $\beta(r)$.

A cosmological void in the asymptotically flat FLRW 
spacetime is characterized by 
$W(r)>1$ near the center and $W(r)\to 1$ in the 
asymptotic region $r\to \infty$. Because of Eq.~(\ref{eq:regularity_1}), 
$W$ must have a maximum which is greater than 1 
in this case. 
Therefore, Eq.~(\ref{eq:shell-crossing_occurrence_W>1}) implies that 
a shell-crossing singularity necessarily appears.
This also applies to PBH formation which has a region where $W>1$.
Therefore, for PBH formation without shell-crossing singularities 
before a future apparent horizon, $W\le 1$ is necessary for all $r\ge 0$.

\subsubsection{Marginally bound shell: $W(r)=1$}

For $W(r)=1$, the shell labeled $r$ begins with a big bang and expands
forever.
For this case, Eq.~(\ref{eq:phi'/phi_W=1}) is relevant. 
$\tilde{X}$ monotonically increases from 0 to $\infty$ as 
time proceeds. 
If $\tilde{a}_{1}(r)>0$, $\psi$ monotonically
increases from $a_{0}(r)$ to $\infty$ and hence there is no
shell-crossing if Eq.~(\ref{eq:regularity_3}) is satisfied. 
This is also the case if $\tilde{a}_{1}(r)=0$, where $\psi$ is constant 
in time.
If $\tilde{a}_{1}(r)<0$, $\psi$ monotonically
decreases from $a_{0}(r)$ to $-\infty$. In this case, 
a shell-crossing singularity appears in the course of 
expansion. 

Therefore, the condition for the occurrence of 
shell-crossing singularity is given by $W'(r)<0$.
If the region inside the 
shell with $W=1$ is unbound ($W>1$), and the region outside 
is bound ($0<W<1$), a shell-crossing singularity appears
on this shell. 

\subsubsection{Criterion in terms of the density distribution}

We can define the following quantities: 
\begin{eqnarray}
\bar{\rho}(t,r)&:=&\frac{M_{,1}(t,r)}{V_{,1}(t,r)}, \\
 \langle \rho \rangle(t,r) &:=&
  \frac{\int_{0}^{r}\bar{\rho}(r',t)W(r')dV(r',t)}{\int_{0}^{r}W(r')dV(r',t)}, \\
M(t,r)&:=&\int^{r}_{0}dr'\int\int dxdy \rho e^{\lambda+2\beta}, \\
V(t,r)&:=&\int_{0}^{r}dr' \int \int dxdy e^{\lambda+2\beta}.
\end{eqnarray}
We can identify $\bar{\rho}$ with the density averaged 
over the two-surface 
$S_{t,r}$, and $\langle \rho \rangle $ with the density averaged 
over the three-ball of 
which the surface is given by $S_{t,r}$. 
The explicit expressions of the above quantities 
for the Szekeres solution are given by
\begin{eqnarray}
 \bar{\rho}(t,r)&=& \frac{S'(r)}{\phi^{2}\phi_{,1}}, \\
 \langle \rho \rangle(t,r)&=& \frac{3S(r)}{\phi^{3}}, \\
 V(t,r)&=& 4\pi \int_{0}^{r}dr'\frac{\phi^{2}\phi_{,1}}{W(r')}, \\
 M(t,r)&=&M(r)=4\pi \int_{0}^{r}dr'\frac{S'(r')}{W(r')}.
\end{eqnarray}

Recall that the regularity at $t=t_{0}$ requires $\beta(r)<1$.
We can find that the ratio of the ball-averaged density to the 
shell-averaged density is calculated to give 
\begin{eqnarray}
 \frac{\langle \rho\rangle
  }{\bar{\rho}}(t,r)=\frac{3S}{S'}\frac{\phi_{,1}}{\phi}=\frac{\psi}{a_{0}(r)}
=\begin{cases}
  \displaystyle\frac{a_{1}(r)}{a_{0}(r)}X+1 & (W(r)\ne 1) \\
  \displaystyle\frac{\tilde{a}_{1}(r)}{a_{0}(r)}\tilde{X}+1 & (W(r)=1)
 \end{cases}
.
\end{eqnarray}
According to the above analysis, we conclude that if there
exists $t$ at which $(\langle \rho \rangle  /\bar{\rho})(t,r)\ge 1$,
then no shell-crossing singularity appears at $r$.
For $0<W(r)< 1$, 
even if  $(\langle \rho \rangle /\bar{\rho})(t,r)< 1$,
no shell-crossing singularity appears before a future 
apparent horizon 
if $(\langle \rho \rangle /\bar{\rho})(t_{\rm FH}(r),r)> \beta(r)$.
For $W(r)>1$, $(\langle \rho \rangle  /\bar{\rho})(t,r)\ge 1$ 
for some $t$ is a sufficient 
condition for the absence of shell-crossing singularity. Even if 
$(\langle \rho \rangle /\bar{\rho})(t,r)< 1$, no 
shell-crossing singularity appears if 
Eq.~(\ref{eq:shell-crossing_occurrence_W>1}) is not satisfied. 
For $W(r)=1$, $(\langle \rho \rangle  /\bar{\rho})(t,r)\ge 1$ 
is a necessary and sufficient condition for the 
absence of shell-crossing singularity.
Note that if  $\bar{\rho}(t,r)$ is a monotonically decreasing function of 
$r$ for fixed $t$, 
$(\langle \rho \rangle /\bar{\rho})(t,r)\ge 1$ holds for all
$r>0$ but not vice versa. 
 
Interestingly, 
Szekeres~\cite{Szekeres:1975dx} has proven that for zero-energy collapse
and time-symmetric collapse, 
the condition for shell-crossing singularities to appear
before the future apparent horizon is the same as we have obtained 
in the simultaneous big bang collapse,  
where the zero-energy collapse means $W(r)=1$ for all $r\ge 0$ and the 
time-symmetric collapse means that $0<W(r)<1$ and $t_{\rm ME}(r)=$const. for
all $r\ge 0$.
The simultaneous big bang condition is a physical condition, which is
independent from the zero-energy 
condition and time-symmetric condition. 
If we assume the simultaneous big bang condition and zero-energy
condition simultaneously, the solution reduces to the flat FLRW solution
and hence there is no perturbation from it. Also, if we assume the 
simultaneous big bang condition and time-symmetric condition simultaneously, 
the solution reduces to the closed FLRW solution and again 
there is no perturbation from it.

\subsection{Shell-focusing singularity}

Let us move on to the possibility that a
shell-focusing singularity is naked.
As we have seen, noncentral shell-focusing singularities are covered by 
a future apparent horizon.
Therefore, we focus on 
central shell-focusing singularity as a result of gravitational
collapse and hence 
we assume $0<W<1$ in the neighborhood of the center. 
In accordance with Eq.~(\ref{eq:regularity_1}), we assume 
\begin{equation}
 1-W^{2}(r)=W_{2}r^{2}+W_{4}r^{4}+O(r^{6}).
\label{eq:W_expansion}
\end{equation}
If $W_{2}=0$, the central shell-focusing singularity does not appear in
finite proper time and hence we assume $W_{2}>0$.
From Eqs.~(\ref{eq:G_-_Y_exp}), 
(\ref{eq:t_FH}), and
(\ref{eq:t_BC}), we find
\begin{eqnarray}
 t_{\rm FH}(r)-t_{\rm BB}&=&\pi
  W_{2}^{-3/2}\left(1-\frac{3}{2}\frac{W_{4}}{W_{2}}r^{2}\right)-\frac{2}{3}r^{3}+O(r^{4}), \\
 t_{\rm BC}(r)-t_{\rm BB}&=&\pi
  W_{2}^{-3/2}\left(1-\frac{3}{2}\frac{W_{4}}{W_{2}}r^{2}\right)+O(r^{4}).
\end{eqnarray} 
Then, if $W_{4}\ge  0$, the central shell becomes
singular only after its neighborhood gets trapped and hence there is 
no null geodesic which emanates from the central singularity.
So, we only have to study the case where $W_{4}<0$.

Equations (\ref{eq:metric}), (\ref{eq:omega}), and (\ref{eq:lambda})
imply that,
along the outgoing radial null tangent vector, the following equation must be
satisfied:
\begin{equation}
 \frac{dt}{dr}=\frac{\phi_{,1}}{W}\left(1-\frac{\phi P_{,1}}{\phi_{,1} P}\right).
\end{equation}
Using Eqs.~(\ref{eq:P_1_nbeta}) and (\ref{eq:psi}),
if there is an outgoing radial null tangent 
from the center, the following 
equation must be satisfied 
\begin{equation}
 \frac{d\ln Y}{d\ln r}=\left(1-\frac{\sqrt{1-W^{2}}}{W}\sqrt{\frac{1}{Y}-1}\right)
\left(\psi+{\vec n}\cdot{\vec \beta}\right)-[a_{1}(r)+a_{0}(r)],
\label{eq:dlnydlnr}
\end{equation}
where we have used Eqs.~(\ref{eq:Friedmann_spherical}) and (\ref{eq:Y}).
Let us put
$Y={\cal Y}r^{\alpha}$
and transform Eq.~(\ref{eq:dlnydlnr}) to 
\begin{equation}
 \frac{d\ln {\cal Y}}{d\ln r}=
\left(1-\frac{\sqrt{1-W^{2}}}{W}\sqrt{\frac{1}{{\cal Y}r^{\alpha}}-1}\right)
\left(\psi+{\vec n}\cdot{\vec \beta}\right)-[a_{1}(r)+a_{0}(r)]
-\alpha,
\label{eq:root_eq}
\end{equation}
where $\alpha$ is constant.
Since $1-W^{2}=O(r^{2})$ for regularity,
we can assume $0\le \alpha \le 2$. If $\alpha>2$, the center 
is trapped and hence there is no outgoing null curve from the
center.
We are interested in the limit to the center, while ${\cal Y}$ approaches 
a positive finite value.
Let us write the right-hand side of Eq.~(\ref{eq:root_eq}) 
as $F(r,{\cal Y})$.

Because of Eq.~(\ref{eq:W_expansion}), 
we can expand $a_{1}(r)$ in the form
\begin{equation}
 a_{1}(r)=-2\frac{W_{4}}{W_{2}}r^{2}+O(r^{4})
\end{equation}
and $\psi$ admits the expansion
\begin{equation}
 \psi\simeq -3\pi\frac{W_{4}}{W_{2}}{\cal Y}^{-3/2}r^{2-\frac{3}{2}\alpha}+1,
\end{equation}
where we have used Eq.~(\ref{eq:expansion_shell-focusing}) and 
left the possible lowest-order terms only.

Note that the smaller the value of  $\alpha$ is, the earlier the
outgoing radial null curve is.
Since we are interested in the earliest one, we first assume $0<\alpha<2$.
Then, the possible lowest-order terms of $F$ are given by 
\begin{equation}
F(r,{\cal Y})\simeq -3\pi\frac{W_{4}}{W_{2}}{\cal Y}^{-3/2}r^{2-3\alpha/2}
-\alpha.
\end{equation}
Let us put $\alpha=4/3$. In this case, we can write $F(r,{\cal Y})$ as 
\begin{equation}
 F(r,{\cal Y})\simeq -3\pi \frac{W_{4}}{W_{2}}{\cal Y}^{-3/2}-\frac{4}{3}.
\end{equation}
Noting that $W_{2}>0$ and $W_{4}<0$, $F(r,{\cal Y})$ can be rewritten as
\begin{equation}
 H(r,{\cal Y})=\frac{4}{3}{\cal Y}_{0}^{3/2}({\cal Y}^{-3/2}-{\cal Y}_{0}^{-3/2}), 
\end{equation}
where 
\begin{equation}
 {\cal Y}_{0}^{3/2}=\frac{9\pi}{4}\frac{(-W_{4})}{W_{2}}.
\end{equation}
Thus, we can see that $F(r,{\cal Y})>0$ if ${\cal Y}<{\cal Y}_{0}$, while 
$F(r,{\cal Y})<0$ if ${\cal Y}>{\cal Y}_{0}$.
This means there exists a single outgoing radial null 
curve which emanates from the central shell-focusing singularity.
For $\alpha\ne 4/3$ but $0<\alpha<2$, we can find that $F$ has no root
for a positive and finite ${\cal Y}$. 
The existence of the radial null tangent ensures the existence of 
an outgoing null geodesic which emanates from the singularity. 
In summary, 
there appears an outgoing null curve with radial null
tangent at the center if and only if $W_{2}>0$ and $W_{4}<0$.
 
The global visibility in the Szekeres spacetime in the sense of 
an event horizon is not analytically tractable in general 
because there is no geodesic which remains radial~\cite{Nolan:2007gy}.
It should again be noted that the global visibility in the sense of an 
event horizon is not physically clearly motivated in the cosmological
context as in the present case.

\section{Models of PBH formation}

\subsection{Condition for PBH formation models}

In the context of cosmological perturbation in the flat FLRW universe,
we cannot generally expect a monotonically decreasing density profile
because the density perturbation should be given randomly 
according to some probability distribution function.
For example, in order to have a model where the deviation from the 
flat FLRW solution falls off sufficiently fast in the asymptotic region, 
the mass excess in the 
overdense perturbation must be compensated by the 
surrounding underdense perturbation. 
This is called a compensated model~\cite{Harada:2015yda}.
Therefore, it is clear that the requirement of a monotonically 
decreasing density profile is too restricted to discuss the 
evolution of nonlinear cosmological perturbation in the Universe.  
Primordial black hole formation in the asymptotically flat FLRW
spacetime is characterized by $0<W<1$ near the center and $W\to 1$
in the asymptotic region $r\to \infty$. 
However, the shell labeled $r$ with $W(r)<1$ eventually collapses.
Therefore, even if $W\to 1$ as $r\to \infty$, the mass of the 
black hole becomes infinite. However, in this situation, $t_{\rm FH}(r)\to
\infty$ as $r\to \infty$ and hence the infinitely massive black hole is 
unphysical. To avoid this unphysical situation, we introduce a cut-off
scale $r_{s}>0$ so that we assume $W(r)=1$ for $r>r_{s}$.

Khlopov and Polnarev~\cite{Khlopov:1980mg}
argue that in the context of the GUT phase transition,  
the occurrence of caustics of massive particles 
will prevent collapse to a black hole 
because modeling by a dust fluid is no longer valid, 
where the matter's pressure gradient force 
cannot be neglected in such a high-density region. 
More precisely, they assume in the LTB solution
that if a central shell-focusing singularity
is not behind a future apparent horizon, then it  
can be regarded to prevent collapse to a black hole because 
the resultant high density implies the breakdown of the dust
approximation of the real matter field and the collapse of the 
shell is impeded by a strong pressure gradient force. We extend 
their assumption to the Szekeres solution.
Also, at shell-crossing singularities, the density grows infinitely
large and hence the dust description should break down.
On one hand, the pressure gradient cannot be neglected there
and the free-fall collapse will be altered. 
On the other hand, it is not so clear whether or not it impedes 
black hole formation.

We can construct a model that is free from
shell-crossing singularity  
if and only if $S$ and $W$ satisfy $0<W(r)\le 1$, $\lim_{r\to \infty}W(r)=1$, 
\begin{equation}
r\left(\ln S^{1/3}\right)'> \beta(r), 
\label{eq:no_shell_crossing_initial}
\end{equation}
and 
\begin{equation}
-r\left[\ln\frac{|1-W^{2}|}{S^{2/3}}\right]'>0.
\label{eq:no_shell_crossing_full}
\end{equation}
We should note that Eq.~(\ref{eq:no_shell_crossing_full}) is equivalent
to
\begin{equation}
 \frac{\langle \rho \rangle}{\bar{\rho}}>1.
\label{eq:ball-average>shell-average}
\end{equation}
It is interesting that once we assume the first condition 
(\ref{eq:no_shell_crossing_initial}), 
no dependence on nonsphericity appears
in the second condition 
even though the model can be highly nonspherical.
However, shell-crossing singularities formed after the 
formation of a future apparent horizon do not affect the formation
of primordial black holes. This means that in the present context
we can relax the condition to that for the absence of shell-crossing 
singularities before the future apparent horizon formation, 
so that Eq.~(\ref{eq:no_shell_crossing_full}) is replaced by 
\begin{equation}
 -r\left[\ln\frac{|1-W^{2}|}{S^{2/3}}\right]'X_{-}(1-W^{2})+r\left(\ln
    S^{1/3}\right)'> \beta(r).
\label{eq:no_shell_crossing_second}
\end{equation}
Clearly, this includes the condition~(\ref{eq:no_shell_crossing_full}). 
In summary, if Eq.~(\ref{eq:no_shell_crossing_full}) or equivalently
(\ref{eq:ball-average>shell-average}) is satisfied, 
no shell-crossing singularity appears from regular initial data
assured by Eq.~(\ref{eq:no_shell_crossing_initial}), while if 
Eq.~(\ref{eq:no_shell_crossing_full}) is not satisfied, 
shell-crossing singularity appears from regular initial data 
assured by Eq.~(\ref{eq:no_shell_crossing_initial}) only after 
a future apparent horizon formation if 
Eq.~(\ref{eq:no_shell_crossing_second}) is satisfied.

\subsection{Global visibility}
In the previous section, we have studied whether or not 
shell-crossing and shell-focusing singularities at $r$ are 
formed before the shell labeled $r$ is trapped by a future apparent
horizon. 
This corresponds to the question whether or not 
singularities are locally naked. 
We can also ask whether or not the singularity is globally naked.
For a singularity, if a light ray which emanates from the 
singularity can get out of the black hole, we call this singularity 
globally naked. If there is no such light ray, then we call this 
singularity globally covered.
In asymptotically flat spacetimes, we can define globally naked
singularities in terms of future null infinities.
In the cosmological setting this is not so well motivated,  
as we have already seen.
Here we deal with this problem by specifying the mass of the black hole.
If we fix the mass of the black hole to $M$, we can identify $r_{s}$ which
gives the black hole mass, i.e., $S(r_{s})=2M$.
To investigate the global visibility, we need to track null geodesics
which emanate from the singularity. However, this problem is very 
difficult to tackle not only because 
null geodesic equations cannot be integrated analytically in general
but also because null geodesics cannot be kept radial even though they
started as radial null ones momentarily.
For this reason, we study a sufficient condition for the singularity 
to be globally covered. This can be done by requiring that the
singularity which occurs at $t$ 
is surrounded by a trapped sphere at the same time $t$.
For the central shell-focusing singularity, this condition yields
\begin{equation}
t_{\rm BC}(0)\ge \min_{0\le x\le r_{s}} t_{\rm FH}(x).
\label{eq:globally_covered_shell-focusing}
\end{equation}
For the shell-crossing singularity at $r$, this condition yields
\begin{equation}
t_{\rm SC}(r)\ge \min_{r\le  x\le  r_{s}}t_{\rm FH}(x).
\label{eq:globally_covered_shell-crossing}
\end{equation}
Equation~(\ref{eq:globally_covered_shell-focusing}) is equivalent to
\begin{equation}
 \lim_{r\to 0}\pi\frac{S(r)}{[1-W^{2}(r)]^{3/2}}\ge \min_{0\le x\le r_{s}}S(x)G_{-}(1-W^{2}(x)),
\end{equation}
while 
Eq.~(\ref{eq:globally_covered_shell-crossing}) has no compact
expression in terms of the arbitrary functions.
The left-hand side of Eq.~(\ref{eq:globally_covered_shell-crossing}) 
decreases if $\beta(r)$ increases, while the right-hand side does not
depend on $\beta(r)$.

\subsection{Exact models of PBH formation}

Here we construct exact models of PBH formation. 
We fix the scaling of $r$ so that $S(r)=r^{3}$ in
this subsection.
Then, only the choice of $W(r)$ determines the reference LTB solution.

\subsubsection{Model A}
\begin{figure}[htbp]
 \begin{center}
  \includegraphics[width=0.5\textwidth]{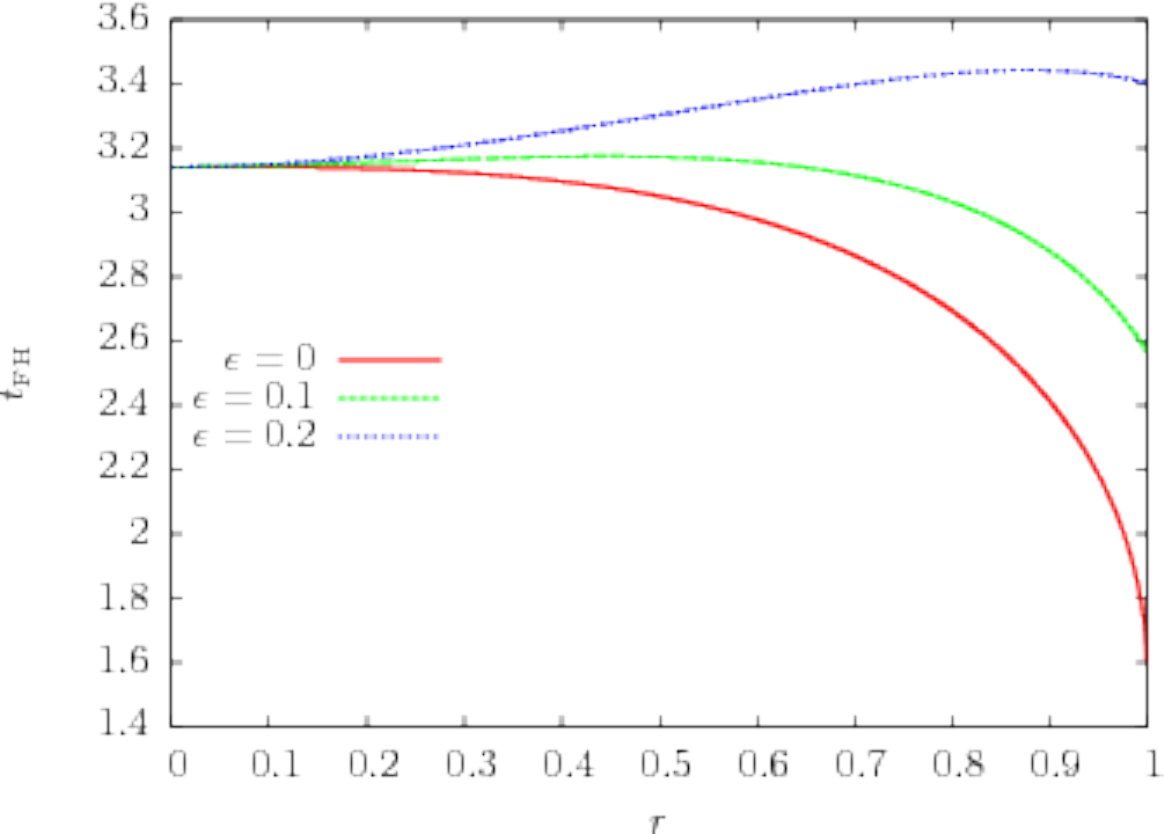}
\caption{\label{fig:apparent_horizon}
The time of future apparent horizon, $t_{\rm FH}(r)$, is
  plotted as a function of $r$. The curves are labeled with 
  the value of $\epsilon$.
  The central shell focusing singularity is globally
  covered for $\epsilon=0$ and $0.1$, although it is locally naked
  for $\epsilon=0.1$.}
 \end{center}
\end{figure}

First, we consider the following natural choice of $W$:
\begin{eqnarray}
 1-W^{2}=\left\{\begin{array}{cc}
  \left(\frac{r}{r_{c}}\right)^{2}\left[1-\epsilon 
   \left(\frac{r}{r_{c}}\right)^{2}\right] & (0<r<r_{c})\\
  0 & (r\ge r_{c})
\end{array}\right. ,
\end{eqnarray}
where $r_{c}$ is a positive constant and $\epsilon$ is constant.
We call this choice of $W(r)$ ``model A''.
For consistency, we need $\epsilon\le 1$.
In this model, as we have seen, 
the central shell-focusing singularity is naked if $\epsilon>0$, 
while it is covered if $\epsilon\le 0$. However, the global visibility 
is not so trivial. Let us choose $r_{c}=1$ and $\epsilon=0$, $0.1$, and $0.2$.
Then, the time of future apparent horizon, $t_{\rm FH}(r)$, is plotted 
in Fig.~\ref{fig:apparent_horizon}. 
We can see that the model with $\epsilon=0.1$ has a central
shell-focusing singularity which is locally naked but globally covered.
For $\epsilon\ge 0$,
since $|1-W^{2}|/S^{2/3}$ is a monotonically decreasing
function of $r$, no shell-crossing singularity appears.
It should be noted that the function $W(r)$ is discontinuous at $r=r_{c}$.
With the above choice, the model is only perturbed within the comoving radius
$r_{s}=r_{c}$ and the region outside it is identical to the flat FLRW universe.
This is also the case irrespective of the choice of the nonspherical
functions because with the simultaneous big bang condition, the 
function $\phi$ becomes identical to that in the flat FLRW spacetime, 
and this reduces to the flat FLRW spacetime 
as we have already seen.
So, in this model the compensation is perfectly realized at $r_{s}=r_{c}$.  

\begin{figure}[htbp]
 \begin{center}
  \includegraphics[width=0.5\textwidth]{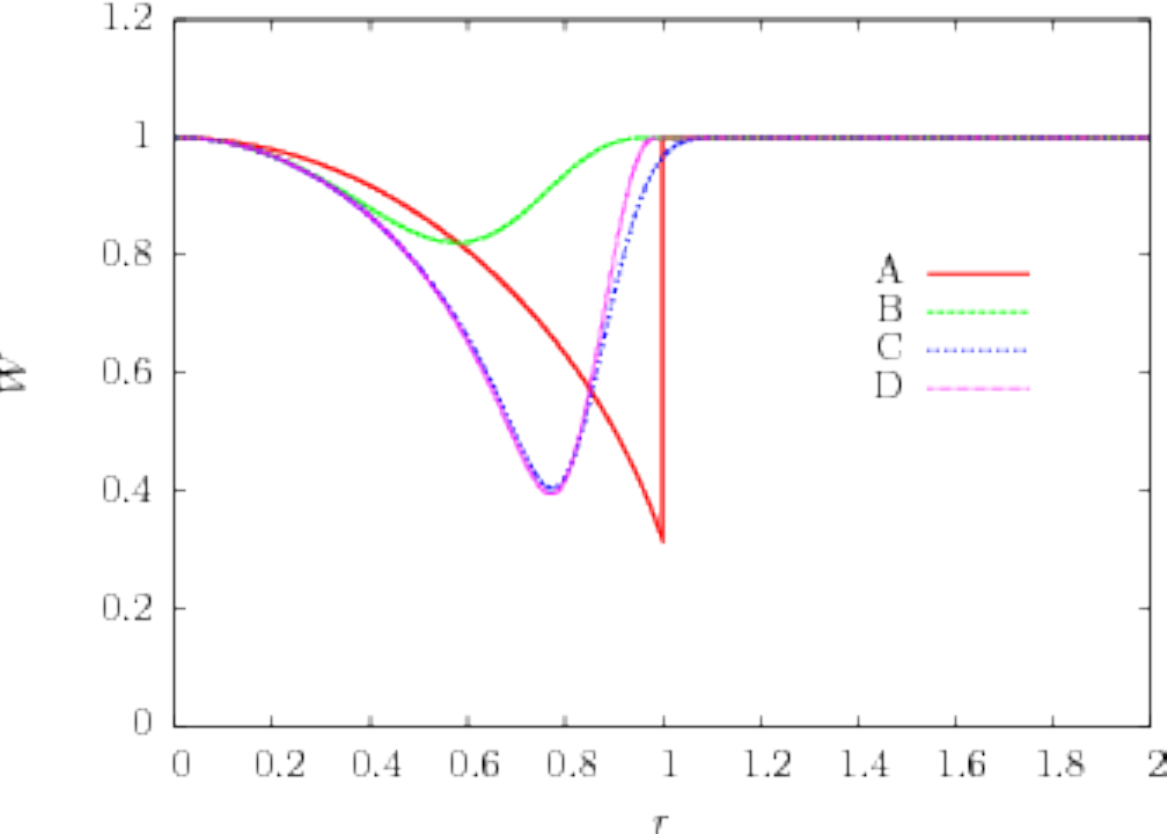}
\caption{\label{fg:energy_function_generator} 
The energy function $W(r)$ for the spherical and nonspherical 
models of PBH formation. The curves labeled with A, B, C, and D
denote the energy functions for models A with $r_{c}=1$ and
  $\epsilon=0.1$, B with $r_{c}=0.8$, $r_{w}=1$, 
C with $r_{c}=0.8$, $r_{w}=0.6$, and $r_{f}=0.3$, and 
D with $r_{c}=0.8$, $r_{w}=1$, $n_{1}=8$, and $n_{2}=10$,
respectively.}
 \end{center}
\end{figure}

\subsubsection{Model B}
Next, we present another example free from shell-crossing
singularity and shell-focusing naked singularity as follows:
\begin{eqnarray}
 1-W^{2}=\left\{\begin{array}{cc}
  \left(\frac{r}{r_{c}}\right)^{2}\left[1-\left(\frac{r}{r_{w}}\right)^{4}\right]^{4} & (0<r<r_{w})\\
  0 & (r\ge r_{w})
\end{array}\right.
,
\end{eqnarray}
where $r_{c}$ and $r_{w}$ are positive constants, and 
the following inequality must be satisfied for $W^{2}> 0$:
$r_{c}/r_{w}>{2^{6}}/{3^{9/2}}$.
We call this choice of function $W(r)$ ``model B''.
In this case, the metric function is $C^{2-}$ and 
we have neither naked shell-focusing singularity 
nor shell-crossing singularity because $W_{4}=0$ 
and $|1-W^{2}|/S^{2/3}$ is a monotonically decreasing function of $r$. 
Also, this model is compensated at $r_{s}=r_{w}$, irrespective of the 
choice of the nonspherical functions.

\subsubsection{Model C}
We can consider another example, for which
\begin{eqnarray}
 1-W^{2}=\left\{\begin{array}{cc}
  \left(\frac{r}{r_{c}}\right)^{2} & (0<r<r_{w})\\
  \left(\frac{r}{r_{c}}\right)^{2}\exp\left[-\left(\frac{r-r_{w}}{r_{f}}\right)^{4}\right] & (r\ge r_{w})
\end{array}\right.
,
\end{eqnarray}
where $r_{c}$, $r_{w}(<r_{c})$, and $r_{f}$ are positive constants, and 
the following inequality must be satisfied for $W^{2}>  0$:
\begin{equation}
\max_{r>r_{w}}\left\{\left(\frac{r}{r_{c}}\right)^{2}
\exp\left[-\left(\frac{r-r_{w}}{r_{f}}\right)^{4}\right]\right\}<1.
\end{equation}
We call this choice of $W(r)$ ``model C''.
This model is also free from shell-focusing naked singularity and 
shell-crossing singularity.
For $0<r<r_{w}$, from the simultaneous big bang condition, 
the function $\phi$ is identical to that in 
the closed FLRW spacetime.
Therefore, the region inside $r_{w}$ is identical to the closed FLRW spacetime 
irrespective of the choice of the nonspherical functions.
Since the fall-off of the function in the asymptotic region $r\to
\infty$ is sufficiently fast, this model is compensated 
at infinity~\cite{Harada:2015yda}, while the mass of the black hole 
becomes infinite because $r_{s}=\infty$.
We should note that the above choices of the energy function are similar
to those presented in Ref.~\cite{Harada:2001kc} for the PBH formation
with the LTB solution but with a slightly different scaling of $r$. 

\subsubsection{Model D}
\begin{figure}[htbp]
 \begin{center}
  \includegraphics[width=0.5\textwidth]{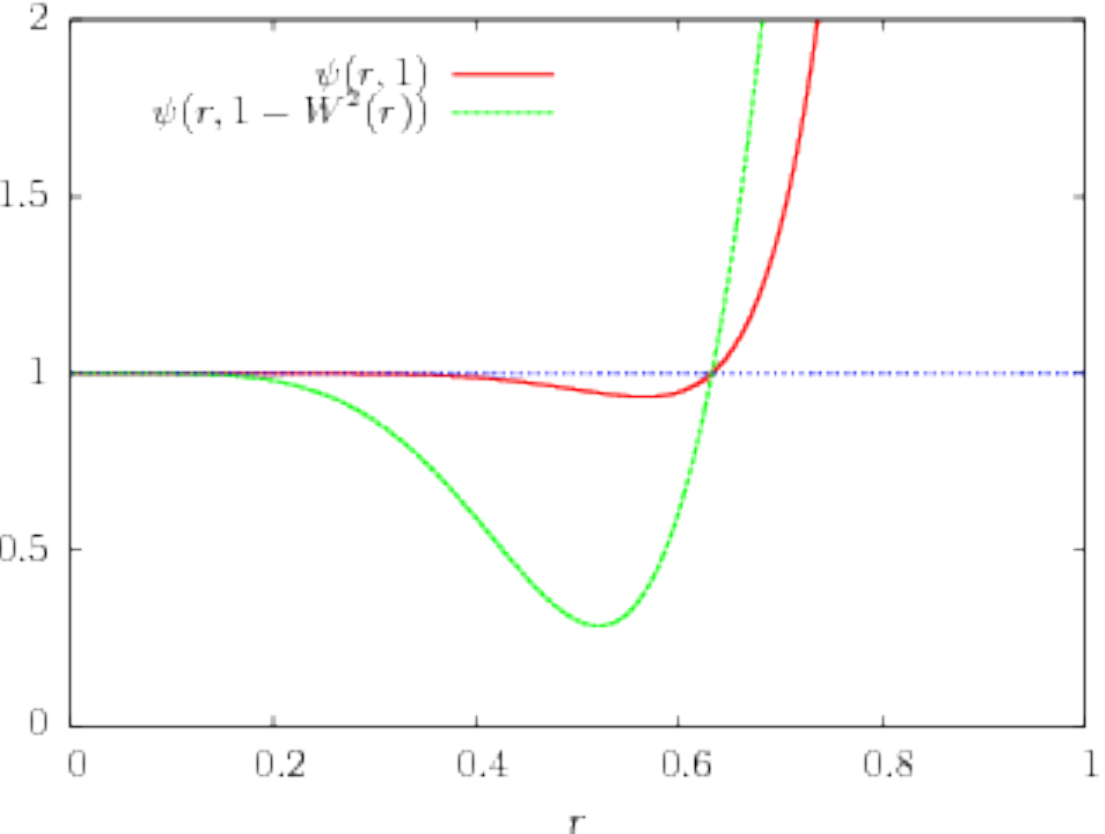}
\caption{\label{fg:no_shell_crossing_second} The functions 
$\psi(r,1)$ and $\psi(r,1-W^{2}(r))$ are plotted as functions of $r$ 
for the polynomial model (model D) 
given by Eq.~(\ref{eq:polynomial_model}), where 
$n_{1}=8$, $n_{2}=10$, $r_{c}=0.8$, and $r_{w}=1$ are adopted.}
 \end{center}
\end{figure}

Now, we propose an example in which shell-crossing appears
but only after the future apparent horizon forms.
Here, we consider a polynomial function of 
$W(r)$ parametrized as follows:
\begin{eqnarray}
 1-W^{2}=\left\{\begin{array}{cc}
  \left(\frac{r}{r_{c}}\right)^{2}\left[1+\left(\frac{r}{r_{w}}\right)^{n_{1}}-2\left(\frac{r}{r_{w}}\right)^{n_{2}}\right]^{4} & (0<r<r_{w})\\
  0 & (r\ge r_{w})
\end{array}\right.
,
\label{eq:polynomial_model}
\end{eqnarray}
where $r_{c}$ and $r_{w}$ are positive constants
and $n_{1}$ and $n_{2}$ are even integers, 
and for consistency, the following inequality must be satisfied:
${r_{c}}/{r_{w}}>\sqrt{f_{max}}$,
where $f_{\max}$ is the maximum value of the function
$f(x)=x^{2}(1+x^{n_{1}}-2x^{n_{2}})^{4}$ for $0<x<1$.
If $n_{1}>2$ and $n_{2}>2$, shell-focusing singularity is 
censored.
We call this choice of $W(r)$ ``model D''.
This model is compensated at $r_{s}=r_{w}$, irrespective of the 
choice of the nonspherical functions.
We put $n_{1}=8$, $n_{2}=10$, $r_{c}=0.8$, and $r_{w}=1$
and plot $\psi(r,1)$ and $\psi(r,1-W^{2})$ as functions of $r$
in Fig.~\ref{fg:no_shell_crossing_second}.

In this case, since $a_{1}(r)$ is not positive definite,
$(\langle \rho \rangle  /\bar{\rho})(t,r)\ge 1$ does not hold for some $r$. 
Noting that $\psi \le \beta(r)$ implies shell-crossing singularities, 
we can understand the evolution of this model.
Notice that $\psi(r,0)=1$, 
$\psi(r,1)$, and $\psi(r,1-W^{2})$ are the values of $\psi$
at the big bang, maximum expansion and future apparent horizon,
respectively.
$\psi$ is a monotonically increasing function of time if $\psi(r,1)>1$
and monotonically decreasing function of time if $\psi(r,1)<1$ 
because $\psi(r,Y)=a_{1}(r)X(Y)+1$. 
We should note that in our scaling, $\beta(r)<1$ must be satisfied
from Eq.~(\ref{eq:regularity_3}). 
So, if $\beta(r)=0$, which is the LTB model, no 
shell-crossing singularity appears before the future apparent horizon 
forms. This is also the case 
if $\beta(r)$ is sufficiently small for $0<r<r_{w}$.
But if there is an $r$ at which
$\psi(r,1-W^{2}(r))<\beta (r)<1$, shell-crossing
singularity appears before the formation of a future apparent horizon. 
Moreover, if there is an $r$ at which 
$\psi(r,1)<\beta(r)<1$, shell-crossing singularity appears 
at $r$ even before the maximum expansion.
This shows that a large dipole moment can promote shell-crossing
singularity occurrence before the formation of an apparent horizon:
if shell-crossing singularity is 
to occur, the larger the dipole moment is and the earlier
the time of shell-crossing singularity becomes.

\section{Conclusion}

We have constructed exact models of spherical and nonspherical 
formation of primordial black holes with the LTB solution and Szekeres's
quasispherical solution of the Einstein equation.
The matter content is restricted to being a pressureless fluid or dust.
The LTB solution contains three arbitrary functions of one variable
with one scaling freedom remaining and hence describes a general 
spherically symmetric and inhomogeneous dust spacetime.  
The Szekeres solution additionally contains four arbitrary 
functions of one variable with one algebraic constraint equation. 
We interpret the Szekeres solution as the nonspherical 
deformation of the LTB solution by adding dipole moment 
distribution. 
These solutions may contain both 
shell-crossing and shell-focusing singularities.
We use these solutions to model the evolution of cosmological 
nonlinear fluctuations. In this context, these singularities are regarded as
the breakdown of the model with a pressureless fluid, where  
strong pressure gradient force instead is at work.

Based on the Szekeres solution, we have analyzed 
how inhomogeneous effects and nonspherical effects
affect the formation of naked singularities 
from cosmological fluctuations.
If the perturbation has a sufficiently homogeneous central region 
and a sufficiently sharp transition in the matching with the background flat FLRW universe, then the central shell-focusing singularity is globally covered. 
Moreover, if the central density concentration is sufficiently large, 
no shell-crossing singularity appears from regular initial 
data, irrespective of the dipole moment distribution.  However, 
if the central density concentration is not sufficiently large,
shell-crossing singularity appears. In this case, 
large dipole moment distribution significantly advances 
shell-crossing singularities and they tend to appear 
before a future apparent horizon.

This is the first exact approach to nonspherical effects 
on primordial black hole formation in full general relativity. 
The current analysis on 
the Szekeres solution is important in the cosmological epoch 
when the effective equation of state is very soft, which
can be realized in the ending phase of inflation.

\section*{Acknowledgments}

The authors are grateful to IUCAA for its hospitality.
TH would like to thank B.~J.~Carr, T.~Kobayashi, Y.~Koga, T.~Nakama, and 
C.~M.~Yoo for fruitful discussions.
This work was supported by JSPS KAKENHI Grant Number 26400282 (TH)
and ISRO-RESPOND grant (ISRO/RES/2/384/2014-2015; SJ).

\appendix

\section{Derivation of Szekeres's quasispherical dust solution\label{sec:derivation_szekeres}}

For this article to be self-contained, 
we present here the derivation of Szekeres's quasispherical dust
solution according to Szekeres~\cite{Szekeres:1974ct} 
and Sects. 19.5.1 and 19.5.2 of Plebanski and 
Krasinski~\cite{Plebanski:2006sd} (amending a few typos there).

We assume the following form of the line element:
\begin{equation}
 ds^{2}=-dt^{2}+e^{2\lambda}dr^{2}+e^{2\omega}(dx^{2}+dy^{2}),
\end{equation}
where $\lambda=\lambda(t,r,x,y)$ and $\omega=\omega(t,r,x,y)$.
The tetrad components with tetrad one-form basis ${\bf
e}^{\hat{0}}:={\bf d}t$, ${\bf e}^{\hat{1}}:=e^{\lambda}{\bf d}r$, 
${\bf e}^{\hat{2}}:=e^{\omega}{\bf d}x$, and ${\bf
e}^{\hat{3}}:=e^{\omega}{\bf d}y$
of the Einstein tensors are calculated to give
\begin{eqnarray}
 G_{\hat{0}\hat{0}}&=&e^{-2\omega}(-\lambda_{,2}^{2}-\lambda_{,3}^{2}-\lambda_{,22}-\omega_{,22}-\lambda_{,33}-\omega_{,33})+e^{-2\lambda}(-3\omega_{,1}^{2}-2\omega_{,11}+2\lambda_{,1}\omega_{,1})\nonumber
  \\
 & &+2\lambda_{,0}\omega_{,0}+\omega_{,0}^{2}, \\
 G_{\hat{0}\hat{1}}&=&-2e^{-\omega}(e^{-\lambda+\omega}\omega_{,1})_{,0}, 
\label{eq:G01}\\
 G_{\hat{1}\hat{1}}&=&e^{-2\lambda}\omega_{,1}^{2}+e^{-2\omega}(\omega_{,22}+\omega_{,33})-3\omega_{,0}^{2}-2\omega_{,00}, 
\label{eq:G11}\\
 G_{\hat{1}\hat{2}}&=&-e^{-\omega}(e^{-\lambda}\omega_{,1})_{,2}, \\
 G_{\hat{1}\hat{3}}&=&-e^{-\omega}(e^{-\lambda}\omega_{,1})_{,3}, \\
 G_{\hat{0}\hat{2}}&=&e^{-\omega}(-\lambda_{,02}-\omega_{,02}-\lambda_{,0}\lambda_{,2}+\omega_{,0}\lambda_{,2}), \\
 G_{\hat{0}\hat{3}}&=&e^{-\omega}(-\lambda_{,03}-\omega_{,03}-\lambda_{,0}\lambda_{,3}+\omega_{,0}\lambda_{,3}), \\
 G_{\hat{2}\hat{2}}&=&e^{-2\omega}(\lambda_{,3}^{2}+\lambda_{,2}\omega_{,2}+\lambda_{,33}-\lambda_{,3}\omega_{,3})+e^{-2\lambda}(\omega_{,1}^{2}+\omega_{,11}-\lambda_{,1}\omega_{,1})
\nonumber \\
& &-\lambda_{,0}\omega_{,0}-\lambda_{,0}^{2}-\omega_{,0}^{2}-\lambda_{,00}-\omega_{,00}, \\
 G_{\hat{3}\hat{3}}&=&e^{-2\omega}(\lambda_{,2}^{2}+\lambda_{,3}\omega_{,3}+\lambda_{,22}-\lambda_{,2}\omega_{,2})+e^{-2\lambda}(\omega_{,1}^{2}+\omega_{,11}-\lambda_{,1}\omega_{,1})
\nonumber \\
& &-\lambda_{,0}\omega_{,0}-\lambda_{,0}^{2}-\omega_{,0}^{2}-\lambda_{,00}-\omega_{,00}, \\
 G_{\hat{2}\hat{3}}&=&e^{-2\omega}(-\lambda_{,23}-\lambda_{,2}\lambda_{,3}+\lambda_{,2}\omega_{,3}+\omega_{,2}\lambda_{,3}).
\end{eqnarray}
Introducing the 
complex coordinates $\xi=x+iy$ and $\bar{\xi}=x-iy$, we find 
\begin{eqnarray}
 G_{\hat{0}\hat{0}}&=&-4
  e^{-2\omega}(\lambda_{,\xi}\lambda_{,\bar{\xi}}+\lambda_{,\xi\bar{\xi}}+\omega_{,\xi\bar{\xi}})+e^{-2\lambda}(-3\omega_{,1}^{2}-2\omega_{,11}+2\lambda_{,1}\omega_{,1})\nonumber
  \\
&&+2\lambda_{,0}\omega_{,0}+\omega_{,0}^{2}, \\
 G_{\hat{0}\hat{2}}-iG_{\hat{0}\hat{3}}&=&2e^{-\omega}(-\lambda_{,0\xi}-\omega_{,0\xi}-\lambda_{,0}\lambda_{,\xi}+\omega_{,0}\lambda_{,\xi}), 
\label{eq:G02-G03}\\
 G_{\hat{0}\hat{2}}+iG_{\hat{0}\hat{3}}&=&2e^{-\omega}(-\lambda_{,0\bar{\xi}}-\omega_{,0\bar{\xi}}-\lambda_{,0}\lambda_{,\bar{\xi}}+\omega_{,0}\lambda_{,\bar{\xi}}), \\
 G_{\hat{2}\hat{2}}-G_{\hat{3}\hat{3}}+2iG_{\hat{2}\hat{3}}&=&
-4e^{-2\omega}(\lambda_{,\xi\xi}+\lambda_{,\xi}^{2}-2\lambda_{,\xi}\omega_{,\xi}), \\
 G_{\hat{2}\hat{2}}-G_{\hat{3}\hat{3}}-2iG_{\hat{2}\hat{3}}&=&
-4e^{-2\omega}(\lambda_{,\bar{\xi}\bar{\xi}}+\lambda_{,\bar{\xi}}^{2}-2\lambda_{,\bar{\xi}}\omega_{,\bar{\xi}}), \\
G_{\hat{2}\hat{2}}+G_{\hat{3}\hat{3}}&=&
4e^{-2\omega}(\lambda_{,\xi\bar{\xi}}+\lambda_{,\xi}\lambda_{,\bar{\xi}})
+2e^{-2\lambda}(\omega_{,1}^{2}+\omega_{,11}-\lambda_{,1}\omega_{,1}) \nonumber \\
& &-2(\lambda_{,0}\omega_{,0}+\lambda_{,0}^{2}+\omega_{,0}^{2}+\lambda_{,00}+\omega_{,00}), 
\label{eq:G22+G33} \\
G_{\hat{1}\hat{2}}-iG_{\hat{1}\hat{3}}&=&-2e^{-\omega}(e^{-\lambda}\omega_{,1})_{,\xi}, \\
G_{\hat{1}\hat{2}}+iG_{\hat{1}\hat{3}}&=&-2e^{-\omega}(e^{-\lambda}\omega_{,1})_{,\bar{\xi}}.
\end{eqnarray}
We assume a dust fluid which moves along the world line of constant $r$,
$x$, and $y$ so that
$
 T^{\mu\nu}=\epsilon u^{\mu}u^{\nu},
$
where 
$
 u^{\mu}=\delta^{\mu}_{0}.
$
We will solve the Einstein equation
$
 G_{\hat{\mu}\hat{\nu}}=8\pi T_{\hat{\mu}\hat{\nu}}.
$

$G_{\hat{1}\hat{2}}=G_{\hat{1}\hat{3}}=0$ imply 
\begin{equation}
 e^{-\lambda}\omega_{,1}=u(t,r),
\end{equation}
where $u(t,r)$ is an arbitrary function. Substituting this into
$G_{\hat{0}\hat{1}}=0$ with Eq.~(\ref{eq:G01}) implies
\begin{equation}
 e^{\omega}u(t,r)=e^{\nu(r,x,y)}.
\end{equation}
We assume $\omega_{,1}\ne 0$
\footnote{
If we instead assume $\omega_{,1}=0$, we obtain another class
of dust solutions, which includes the Datt-Ruban family of solutions in
the case of spherical symmetry. See Ref.~\cite{Plebanski:2006sd} for details.}
Putting $\phi(t,r):=1/u(t,r)$, we
find 
\begin{eqnarray}
e^{\lambda}&=&h(r)\phi(t,r)\omega_{,1}, \\
e^{\omega}&=&\phi(t,r)e^{\nu(r,x,y)},
\end{eqnarray}
where $h(r)$ appears by rescaling $r$ to $\tilde{r}(r)$. Then,
$G_{\hat{1}\hat{1}}=0$ with Eq.~(\ref{eq:G11}) implies that there exists a function $k(r)$ such that 
\begin{eqnarray}
 e^{-2\nu}(\nu_{,22}+\nu_{,33})+\frac{1}{h^{2}(r)}&=&-k(r), 
\label{eq:G11_separate_1}\\
 2\phi\phi_{,00}+\phi_{,0}^{2}&=&-k(r).
\label{eq:G11_separate_2}
\end{eqnarray}

Since the Ricci scalar $^{(2)}R$ of the 
two-surface $\Omega_{2}$ 
with the line element $d\Omega_{2}^{2}=e^{2\nu}(dx^{2}+dy^{2})$
is given by 
$
 ^{(2)}R=-2e^{-2\nu}(\nu_{,22}+\nu_{,33})
$,
Eq.~(\ref{eq:G11_separate_1}) implies that the two-surface $S_{t,r}$
is a constant curvature space with 
\begin{equation}
 ^{(2)}R=2\left[\frac{1}{h^{2}(r)}+k(r)\right].
\end{equation}
Using the complex coordinates, Eq.~(\ref{eq:G11_separate_1}) implies
$
 (e^{-2\nu}\nu_{,\xi\bar{\xi}})_{,\xi}=0.
$
This can be integrated to give
$
 \nu_{,\xi\xi}-\nu_{,\xi}^{2}=\tau(\xi).
$
We should note that $d\Omega_{2}^{2}=e^{2\nu}d\xi d\bar{\xi}$
in the complex coordinates.
By rescaling 
$\xi=f(\tilde{\xi})$ and $\bar{\xi}=\bar{f}(\tilde{\bar{\xi}})$,
$\nu$ is transformed to $\tilde{\nu}$ such that
$
 e^{2\tilde{\nu}}=e^{2\nu}f'(\tilde{\xi})\bar{f}'(\bar{\tilde{\xi}}).
$
We can choose the function $f$ so that $\tilde{\tau}(\tilde{\xi})=0$.
Then, we redefine $\tilde{\xi}$ as $\xi$ so that 
$
 \nu_{,\xi\xi}-\nu_{,\xi}^{2}=0.
$
This implies 
$
 (e^{-\nu})_{,\xi\xi}=(e^{-\nu})_{,\bar{\xi}\bar{\xi}}=0,
$
since $e^{-\nu}$ is real. This can be integrated to give
\begin{equation}
 e^{-\nu}=P(r,\xi,\bar{\xi}),
\end{equation}
with
\begin{equation}
P(r,\xi,\bar{\xi}):=A(r)\xi\bar{\xi}+B(r)\xi+\bar{B}(r)\bar{\xi}+C(r),
\label{eq:P_xi}
\end{equation}
where $A(r)$ and $C(r)$ are real and $B(r)$ is complex. We can rewrite
this in the following form:
\begin{equation}
 P(r,x,y)=A(r)(x^{2}+y^{2})+2B_{1}(r)x+2B_{2}(r)y+C(r),
\end{equation}
where $B(r)=B_{1}(r)+iB_{2}(r)$ and 
$x$, $y$, $B_{1}(r)$, and $B_{2}(r)$ are real.

Since 
$
 -e^{-2\nu}(\nu_{,22}+\nu_{,33})=A(r)C(r)-B_{1}^{2}(r)-B_{2}^{2}(r)
$, 
Eq.~(\ref{eq:G11_separate_1}) implies 
\begin{equation}
 A(r)C(r)-B(r)\bar{B}(r)=A(r)C(r)-B_{1}^{2}(r)-B_{2}^{2}(r)=\frac{1}{4}g(r),
\end{equation}
where 
\begin{equation}
 g(r):=\frac{1}{h^{2}(r)}+k(r).
\end{equation}

Equation~(\ref{eq:G11_separate_2}) can be integrated to give
\begin{equation}
 \phi_{,0}^{2}=-k(r)+\frac{S(r)}{\phi},
\label{eq:phi,0^2}
\end{equation}
where $S(r)$ is an arbitrary function. 

We can put $\tilde{\phi}(t,r)=f(r)\phi(t,r)$ with an arbitrary
nonvanishing function $f(r)$. If we transform $\phi$ to $\tilde{\phi}$
keeping the metric functions and coordinates unchanged, we should transform 
$e^{\nu}$, $A$, $B_{1}$, $B_{2}$, $C$, $h$, $k$, $S$, and $g$ as follows:
$e^{\tilde{\nu}}=(1/f)e^{\nu}$, $\tilde{A}=fA$, 
$\tilde{B}_{1}=fB_{1}$, $\tilde{B}_{2}=fB_{2}$,
  $\tilde{C}=fC$, $\tilde{h}=(1/f)h$, $\tilde{k}=f^{2}k$, 
$\tilde{S}=f^{3}S$, and $\tilde{g}=f^{2}g$.
If $g\ne 0$, we can make $|\tilde{g}|=1$ by choosing 
$
 f=1/\sqrt{|g|}.
$
Thus, we have
$
 \tilde{g}=K$, where $K:=\mbox{sign}(g)$, 
so that 
\begin{eqnarray}
\tilde{A}(r)\tilde{C}(r)-\tilde{B}_{1}(r)\tilde{B}_{2}(r)=\frac{K}{4}
\end{eqnarray}
and 
\begin{eqnarray}
\frac{1}{\tilde{h}^{2}(r)}+\tilde{k}(r)=K.
\end{eqnarray}
Hereafter we redefine $\tilde{\phi}$
as $\phi$, and so on.
Defining $W(r):=1/h(r)$, we can write $k(r)$ in terms of $W(r)$ as
$
 k(r)=K-W^{2}(r).
$
Thus, Eq.~(\ref{eq:phi,0^2})
becomes
\begin{equation}
 \phi_{,0}^{2}=W^{2}(r)-K+\frac{S(r)}{\phi}.
\label{eq:Friedmann}
\end{equation}
The solutions are called quasispherical, quasiplanar, and quasi-pseudospherical,
respectively, for $K=1$, $0$, and $-1$.
Equation~(\ref{eq:Friedmann})
is integrated to give
\begin{equation}
 \pm \int^{\phi}_{0}\displaystyle\frac{d \varphi}{\sqrt{W^{2}(r)-K+\frac{S(r)}{\varphi}}}=t+H(r),
\label{eq:H}
\end{equation}
where $H(r)$ is an arbitrary function of $r$. Thus, the solution
contains seven arbitrary functions with an algebraic constraint.

Substituting the explicit expressions for $\lambda$ and $\omega$ into 
$G_{\hat{0}\hat{0}}=8\pi\epsilon$, after a rather lengthy
calculation we can obtain the following 
compact expression for $\epsilon$:
\begin{equation}
 8\pi\epsilon=\frac{S'P-3SP_{,1}}{\phi^{2}(\phi_{,1}P-\phi P_{,1})}.
\end{equation}
Substituting the expressions for $\lambda$ and $\omega$, we can show that
Eqs.~(\ref{eq:G02-G03})--(\ref{eq:G22+G33}) identically vanish  and
hence all the components of the Einstein equations are satisfied.

\end{document}